%% file: main.tex
\documentclass[conference,compsoc]{IEEEtran}

\input{preamble}

\usepackage{flushend}

\begin{document}

\title{\codename: Fixing the Android TEE Ecosystem with Arm CCA}

 \author{
\IEEEauthorblockN{Mark Kuhne\quad Supraja Sridhara \quad Andrin Bertschi \quad Nicolas Dutly \quad Srdjan Capkun \quad Shweta Shinde}
\IEEEauthorblockA{ETH Zurich}
}

\maketitle
 
\input{sections/00_abstract}

\IEEEpeerreviewmaketitle
\pagestyle{plain}
\input{sections/01_introduction}
\input{sections/02_tz-limitations}

\input{sections/03_potential-sols}

\input{sections/04_overview}

\input{sections/06_security-analysis}

\input{sections/07_implementation}

\input{sections/08_evaluation}

\input{sections/09_case-studies}

\input{sections/10_related-work}

\input{sections/11_conclusion}
\input{sections/12_acknowledgements}

\bibliographystyle{IEEEtran}
\bibliography{IEEEabrv,references}

\appendices

\input{sections/appendix}
\end{document}

%% file: preamble.tex
\ifCLASSOPTIONcompsoc
  \usepackage[nocompress]{cite}
\else
  \usepackage{cite}
\fi

\ifCLASSINFOpdf
\else
\fi

\usepackage{array}

\usepackage{amsmath}

\usepackage{graphicx}
\usepackage{xcolor}
\usepackage{xspace}
\usepackage{enumitem}
\usepackage{tikz}
\usepackage{hyperref}
\usepackage{url}

\usepackage{breakurl}

\usepackage{booktabs}
\usepackage{multirow}
\usepackage{adjustbox}
\usepackage{colortbl}
\usepackage{cleveref}

\usepackage{siunitx}
\usepackage{listings}

\usepackage{tikz-cd}
\usepackage{amssymb}
\usepackage{adjustbox}
\usepackage{pifont}
\usepackage{longtable}

\newcommand{\codename}{{\sc Aster}\xspace}

\newcommand{\etal}{et al.\xspace}

\newcommand{\gptns}{GPT$_{n}$\xspace}
\newcommand{\gptrs}{GPT$_{rs}$\xspace}

\newcommand{\sservices}{SBSes\xspace}
\newcommand{\Sservices}{SBSes\xspace}
\newcommand{\SServices}{SBSes\xspace}

\newcommand{\SService}{SBS\xspace}
\newcommand{\sservice}{SBS\xspace}

\newcommand{\tsfw}{bootloader\xspace}
\newcommand{\ta}{TA\xspace}
\newcommand{\tf}{TF\xspace}

\crefname{figure}{Figure}{Figure}
\crefname{section}{Section}{Section}
\crefname{appendix}{Appx.}{Appx.}
\crefname{table}{Table}{Table}
\crefname{listing}{Listing}{Listing}

\SetEnumitemKey{myenum}{leftmargin = *}

\newlist{invenum}{enumerate}{1}
\setlist[invenum]{label*=INV\arabic*:~,ref=INV\arabic*}
\makeatletter
\newcommand\myitem[1][]{%
  \if\relax\detokenize{#1}\relax
    \item\relax
  \else
    \protected@edef\@currentlabel{\textcolor{black}{INV$_{#1}$}} %
    \item[\textbf{INV$_{#1}$:}]
  \fi}
\makeatother

\newlist{secenum}{enumerate}{1}
\setlist[secenum]{label*=S\arabic*:~,ref=S\arabic*}
\makeatletter
\newcommand\secitem[1][]{%
  \if\relax\detokenize{#1}\relax
    \item\relax
  \else
    \protected@edef\@currentlabel{\textcolor{black}{S$_{#1}$}} %
    \item[\textbf{S$_{#1}$:}]
  \fi}
\makeatother

\newcommand{\ssecstate}{$\tt{sec\_state}$\xspace}
\newcommand{\loc}{LoC\xspace}
\newcommand{\rsiexaccess}{$\tt{rsi\_ex\_access}$\xspace}
\newcommand{\rsimmio}{$\tt{rsi\_mmio}$\xspace}

\newcommand{\setupbase}{S$_{b}$\xspace}
\newcommand{\setupour}{S$_{a}$\xspace}
\newcommand{\android}{Android\xspace}

\crefname{figure}{Fig.}{Fig.}
\crefname{appendix}{Appx.}{Appx.}
\crefname{table}{Tab.}{Tab.}
\crefname{listing}{Lst.}{Lst.}

%% file: sections/00_abstract.tex
\begin{abstract}

The Android ecosystem relies on either TrustZone (e.g., OP-TEE, QTEE, Trusty) or trusted hypervisors (pKVM, Gunyah) to isolate security-sensitive services from malicious apps and Android bugs. 
TrustZone allows any secure world code to access the  normal world that runs Android. Similarly, a trusted hypervisor has full access to Android running in one VM and security services in other VMs.
In this paper, we motivate the need for mutual isolation, wherein Android, hypervisors, and the secure world are isolated from each other. Then, we propose a 
sandboxed service abstraction, such that a sandboxed execution cannot access any other sandbox, Android, hypervisor, or secure world memory. 
We present \codename which achieves these goals while ensuring that sandboxed execution can still communicate with Android to get inputs and provide outputs securely. 
Our main insight is to leverage the hardware isolation offered by Arm Confidential Computing Architecture (CCA).
However, since CCA does not satisfy our sandboxing and mutual isolation requirements, \codename repurposes its hardware enforcement to meet its goals while addressing challenges such as secure interfaces, virtio, and protection against interrupts. 
We implement \codename to demonstrate its feasibility and assess its compatibility. 
We take three case studies, including one currently deployed on Android phones and insufficiently secured using a trusted hypervisor, to demonstrate that they can be protected by \codename.
\end{abstract}

%% file: sections/01_introduction.tex
\section{Introduction}

Android-based mobile phones consistently own more than 70\% market share~\cite{mobile-os-share-2024}.
From its inception, Android has been conscious of ensuring security-by-design, evident from components that are integral parts of its architecture (e.g., permission system, verified  boot)~\cite{android-security-docs}.
Android's main goal is to protect itself as well as benign apps from potentially malicious apps that the user may install. 
Like any other software that has complexity, large codebase size, and uses memory unsafe languages, Android has been subject to several critical bugs. A malicious application can exploit these bugs to not only attack other applications, but also to subvert Android's protections. This poses a severe risk to user's private information and Android itself. 

Trusted execution has emerged as a promising solution to this problem. 
If the trusted hardware can isolate security-sensitive services and apps (e.g., OTP, updates, key generation) from user applications and Android, then an attacker can no longer exploit bugs to exfiltrate or compromise such services. Since Arm is the most dominant ISA used for mobile phones, most phones are shipped with TrustZone-based trusted execution. 
Intentionally built for simplicity, TrustZone offers a secure world that can house services that need to be protected from apps and Android which run in their own world called the normal world. 

Despite its wide adoption, TrustZone-based solutions---of which there are plenty even in production phones---suffer from a crucial pitfall. By design, the secure world has more privileges than the normal world, meaning any code executing in the secure world can access all the normal world memory, including Android.
On its own, this design choice may not seem alarming. However, over the past decade, several secure world-bound trusted applications and services have suffered from buggy implementations~\cite{sok-tz-vuln}. 
This has spawned a rich line of research, including trusted OSes and hypervisors executing in the secure/normal world to stop the buggy apps from accessing normal world memory. Unfortunately, attackers have found ways to escape this hierarchical protection by either exploiting bugs in secure world OSes and hypervisors or design flaws in their implementations or interfaces. Containing these attacks has been architecturally challenging, since bugs in the secure world may lead to a complete compromise of the phone. In fact, instead of using malicious apps to attack Android, attackers simply exploit bugs in the secure world to attack Android~\cite{boomerang}. Lastly, since the manufacturer controls the secure world, it is nearly impossible for Android or users to employ any protection mechanisms in the secure world. 

In response to these challenges, Android has moved to using normal-world virtualization to house its own security services, in the hopes of abandoning  using the secure world completely~\cite{avf}. At least with Android's normal-world hypervisor, referred to as Android Virtualization Framework (AVF), Google can be assured that its own design and implementations are trustworthy (e.g., Rust based implementation, small TCB, rigorous testing). AVF also allows users to launch other hypervisors if needed (e.g., Qualcomm's Gunyah instead of Google's protected KVM).
Nonetheless, the manufacturers can and do continue to deploy potentially buggy implementations of trusted services and apps in the secure world. AVF simply cannot stop attacks from such services squarely because of inherent ISA decisions. 

In this paper, we take a measured look at the existing Android ecosystem and prior works on attacks and defenses. 
We identify three main gaps: violation of the principle of least privilege, lack of sandboxing, and poorly or completely undefined interfaces.
Based on these observations, we define a new abstraction called {\em Sandboxed Service (SBS)} to capture security-sensitive services that desire to run in the presence of untrusted software such as malicious apps, buggy Android, and  compromised secure world. Next, SBS cannot access normal or secure world memory. Lastly, normal and secure world should not be able to access each other's memory. 
We analyze the complete design space of existing and potential solutions based on TrustZone to achieve our goals, only to conclude with an impossibility result. 

Next, we turn our attention to Realm Management Extension (RME), an Arm extension which enables Arm Confidential Computing Architecture (Arm CCA), for a potential solution~\cite{cca}. 
CCA is primarily built for enabling confidential virtual machines on Arm platforms by supporting 4 worlds, the existing normal and secure world with the addition of a realm world for VMs and a root world for security enforcement. 
At first glance, it may be tempting to conclude that the realm world can be easily used to house SBS. But CCA inherits the TrustZone decision---both secure and realm world can access normal world. 
Thus any solution based on CCA stands to suffer from same challenges as TrustZone, at least for mobile platforms. 
But perhaps not all is lost with CCA. In particular, unlike TrustZone which has a rigid address space controller, CCA uses the notion of programmable Granule Protection Tables (GPTs). Specifically, it allows programming each CPU core with its own GPT. Different GPTs can map the same memory address as belonging to different worlds. This opens up an opportunity to enforce view-based spatial isolation---depending on what code a core is executing (e.g., Android vs SBS), one can use a GPT that denies or allows access to a physical address. 

We present \codename, a novel design that enables the SBS abstraction using Arm CCA. 
First, \codename uses a two-GPT design to achieve mutual world-level isolation i.e., normal and realm worlds cannot access each other. Then, \codename completely stops the secure world from accessing the normal world. Put together \codename achieves ideal world-level isolation desired for Android. 
Next, \codename ensures that any compromise within an SBS is contained.
We leverage stage-two page tables and then enforce a strict communication interface for an SBS to ensure that it cannot escape the sandbox. 
Lastly, \codename addresses several challenges such as hardware attestation, virtual IO, and compatibility.

We implement \codename on a simulator, due to lack of CCA-enabled hardware. Our prototype shows that \codename can easily support Android and existing apps in the normal world. Our minimal changes allow us to integrate \codename into Android Virtualization Framework to launch SBSes. We take existing services that run on Android-spawned VMs and port them to SBSes running in a two-way sandbox in realm world.
We make the following contributions in this paper: 
\begin{itemize}
    \item
    We show that existing as well as potential solutions based on TrustZone are insufficient to achieve mutual isolation and secure sandbox abstraction (SBS). 
    \item
    \codename repurposes Arm CCA GPTs and augments it with critical sandboxing and interfaces to realize SBS.
    \item
    We port 3 cases studies from Android Virtualization Framework to show  \codename's reasonable overheads. 
    
\end{itemize}

%% file: sections/02_tz-limitations.tex
\section{Motivation}
\label{ssec:tz-attacks}

Arm TrustZone divides the system into the normal world and the secure world. 
In each of these worlds, the Arm architecture provides privilege levels known as exception levels (EL) ranging from EL0 to EL3. 
EL3 is the most privileged and is usually where a trusted firmware (\tf) executes and is part of the secure world. 
On platforms with virtualization support, the hypervisor operates at EL2, while the guest kernel and user-mode software run at EL1 and EL0 respectively.
On phones, the secure world runs trusted apps (\ta{s}) which are isolated by a trusted OS, while the normal world executes \android. 
TrustZone tags every memory access with a world bit, which the TrustZone Address Space Controller (TZASC) uses to filter accesses to memory.
The TZASC is configured by the secure world and ensures that the normal world cannot access the secure world. 
However, in TrustZone, the secure world can access all normal world memory making it more privileged.

\noindent{\bf Overprivileged Secure World.} Many previous works have abused the privileges of the secure world to compromise the security of TrustZone-enabled devices. 
These attacks depend on two main observations. 
First, the \ta{s} running in the secure world have unrestricted access to normal world memory; this level of privilege of the \ta{s} is functionally unnecessary. 
Second, \ta{s} have been known to have a wide-array of implementation bugs (e.g., classic input validation errors leading to buffer overflows, bad interface sanitization)~\cite{sok-tz-vuln, ma2023return}.
Put together, these privileged and buggy \ta{s} have opened a large attack surface which an unprivileged \android app can exploit. 
For example, Qualcomm's TrustZone implementation exposes a memory-mapping interface that allows a \ta to map all normal world memory into its address space~\cite{sok-tz-vuln}. 
So, attackers in the normal world (e.g., through attacker-controlled \android apps) who compromise a buggy \ta can gain full access to privileged \android services and kernel to compromise its security (\cref{fig:settings}(a)). 

\noindent{\bf Buggy Interface Definitions and Shared Memory.} Other works show more subtle attacks that use bugs in shared memory implementations of \ta{s} and exploit the fact that the secure world is overprivileged to gain arbitrary read and write to normal-world memory~\cite{boomerang, sok-tz-vuln}. 
For example, Boomerang introduces a class of vulnerabilities that allow malicious unprivileged \android apps to read and write to any normal world memory by exploiting shared buffers with the \ta~\cite{boomerang}. 
Concretely, a malicious \android application passes a pointer to memory it cannot directly access (e.g., \android kernel memory, memory of other applications in the normal world) through the shared buffer. 
The privileged \ta successfully dereferences this pointer and passes the results back to the malicious application. 
\ta{s} can choose to either use standard interfaces (e.g., Global Platform) or introduce their own interfaces~\cite{global-platform, sok-tz-vuln}. 
In either case, these interfaces are prone to developer bugs. 
So, an attacker in the normal world can exploit these interface bugs in a privileged secure world to compromise device security.

\begin{figure*}[]
    \centering
\includegraphics[scale=0.75]{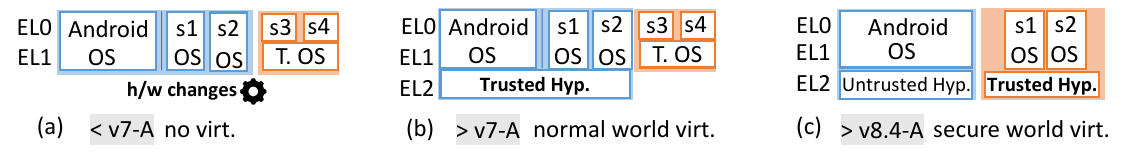} 
         \caption{Security in TrustZone. Blue: Normal world Orange: secure world. s1, s2 are sandboxes. s3, s4 are secure world \ta{s} (a) no virtualization support (b) virtualization support in the normal world (c) virtualization support in the secure world}
    \label{fig:secure-world}
 \end{figure*}
 
\noindent{\bf Lack of Hardware-Based Attestation.} 
Cerdeira~\etal present attacks that arise from the absence of hardware-based attestation primitives in TrustZone~\cite{sok-tz-vuln}.
In the absence of hardware-based attestation, software executing in the secure world creates attestation reports for local and remote attestation. 
If an attacker compromises this software in the secure world using the attacks discussed above, then these attestation primitives are no longer reliable as the attacker can create fake attestation reports.
As a direct consequence of this, validating the version of \ta{s} is unreliable and attackers can downgrade them to known vulnerable versions even if they are patched~\cite{chen2017downgrade, sok-tz-vuln}. 
Therefore, to ensure security, the system must support hardware-based attestation primitives.

\noindent{\bf Insecure \ta Implementations.} Cerdeira~\etal discuss several attacks that stem from  bugs in the trusted OS (e.g., non-standard, expressive, and buggy system call interfaces) and bugs in \ta{s}~\cite{sok-tz-vuln}. 
These attacks rely on bad configurations (e.g., debugging channels left open, lack of ASLR) and implementation bugs where classic security practices (e.g., stack canaries, guard pages) are omitted, or concurrency is incorrectly handled (e.g., race conditions in \ta{s}). 
As an ecosystem, we cannot control the security that individual developers consider when building their \ta{s}. 
Currently, the intra-world isolation between the \ta{s} is performed by a trusted OS in the secure world. 
This trusted OS has a large expressive interface that it exposes to the \ta{s} and is shown to have bugs~\cite{sok-tz-vuln}. 
The trusted OS performs unsafe operations (e.g., directly access the service's memory, map shared libraries into several \ta{s} at the same time) which attackers can abuse to break the intra-world isolation. 
Consequently, a single buggy \ta can put all other \ta{s} in the secure world at risk of being compromised (e.g., leaking keys in the secure world) (\cref{fig:settings}(a)).
Further, because the secure world is more privileged, attackers can use the buggy \ta to escalate privilege and break device security.

\noindent{\bf Can TrustZone Platforms Solve These Problems?}
Previous works have explored different approaches to solve the security issues in TrustZone. 
Several works have proposed designs that build sandboxes that contain the effects of buggy \ta{s}. 
Before Arm introduced virtualization support, previous works proposed designs that change the hardware to ensure that the sandboxes are isolated from each other in the normal world (\cref{fig:secure-world}(a))~\cite{brasser2019sanctuary}. 
Other works propose building sandboxes in the secure world, either by using a trusted hypervisor in secure world EL1 or by using special hardware on some TrustZone platforms~\cite{li2019teev, cerdeira2022rezone}. 
Then, Armv7-A introduced virtualization support in the normal world.  
On Armv7-A, works explore designs which sandbox \ta{s} in the normal world using a trusted hypervisor (\cref{fig:secure-world}(b))~\cite{hua2017vtz}.
\android has recently adopted this approach and introduced a trusted hypervisor that executes in the normal world and creates sandboxes called protected VMs~\cite{avf, gunyah}. 
Later, Armv8.4-A introduced virtualization support in the secure world.  
Using this, previous works build designs that employ a trusted hypervisor to create sandboxes in the secure world (\cref{fig:secure-world}(c))~\cite{li2021twinvisor, hafnium, kwon2019pros}. 

While these approaches create sandboxes for the \ta{s}, they do not consider enforcing a notion of least privilege in the secure world i.e., the secure world should only have selective access to the memory it needs, instead of an access to the entire normal world by default.
Without such a notion, while the sandboxes isolate the \ta{s} from each other using translation tables managed by the secure world (OS or hypervisor), attackers can still exploit their bugs to access the normal world and compromise device security (\cref{fig:settings}(b)). 
With just TrustZone, it is not possible to implement a policy of least privilege for the secure world as the TZASC that performs the memory access control is always configured by the secure world~\cite{shelter}.

\section{Problem Formulation}
\label{ssec:problem-formulation}
\cref{ssec:tz-attacks} highlights the need for strong isolation between the normal and secure worlds and calls for introducing a notion of least privilege for the secure world. 
If we ensure strong mutual isolation between the \ta{s} and \android, and limit the memory accessible to the \ta{s}, this eliminates the attacks that exploit bugs in the \ta{s} to compromise the normal world. 
For example, the attacks in~\cref{ssec:tz-attacks} that use \ta{s} to map all normal world memory or de-reference pointers to normal world memory will be thwarted. 

However, just de-privileging the \ta{s} and ensuring mutual isolation between the normal and secure worlds is not sufficient. 
Even if the compromised \ta{s} cannot access normal world memory, attackers can still use them to mount attacks on other \ta{s} on the trusted OS. 
To prevent such cross \ta attacks, we need to ensure that the \ta{s} are further sandboxed. 
Sandboxing \ta{s} would ensure that even if an attacker compromises a \ta, its effects are contained to that service. 
We refer to such sandboxed \ta{s} as \emph{sandboxed services} (\sservices). 
A sandboxed service is any trusted computation that is used by normal world \android apps. 
\Sservices can be bare-metal binaries, binaries with thin library OSes, or full VMs. 

Mutual isolation and sandboxing ensure that the bugs in the \sservices are contained to their sandbox. 
From our analysis in~\cref{ssec:tz-attacks}, we see that we can improve the security of \sservices by narrowing the interfaces they use. 
For example, ensuring that shared memory between the \sservices and normal world is non-executable would stop attacks where an attacker executes malicious code from this interface in the \sservice. 
As shown by previous works, we can reduce the interface attack surface by introducing global settings for interface security (e.g., non-executable shared memory)~\cite{elasticlave}. 
Further, by using standard interfaces we can use a rich ecosystem of interface testing and validation tools (e.g., fuzzers, sanitizers)~\cite{feng2016bindercracker, feng2016understanding, liu2020fans}.

Finally, \sservices should be deployed on a platform that supports hardware-based local and remote attestation primitives that cannot be compromised by a software adversary.
Using these primitives, trusted software can validate (e.g., versions, signatures) the \sservices before deploying them and remote verifiers can check  attestation reports before transferring secrets to the \sservices (e.g., code for trusted compilation). 
In summary, to better protect the  applications and services while retaining functionality, mobile platforms need to achieve the following security primitives:
\begin{secenum}[labelindent=0pt,labelwidth=0.75em,leftmargin=!, wide=0pt]
    \secitem[sand]\label{s:sand} {\em Sandboxing.} During the \sservices execution, any effects of buggy \sservices are contained within the sandbox.
    \secitem[iso]\label{s:iso} {\em Mutual Isolation.} The \sservices, \android, and secure world should be mutually isolated by the hardware. 
      \secitem[int]\label{s:int} {\em Secure Interfaces.} A \sservice should use secure interfaces to communicate with \android and other \sservices.
    \secitem[att]\label{s:att} {\em Attestation.} All \sservices should be attested before launch using hardware-based attestation. 
\end{secenum}

%% file: sections/03_potential-sols.tex
\section{Challenges in Using Arm CCA}
We aim to design a platform that can be used by \android applications to deploy \sservices while ensuring our security primitives. We investigate if Arm CCA provides the mechanisms to deploy \sservices on \android.

\begin{figure}
    \centering
     \includegraphics[scale=0.55]{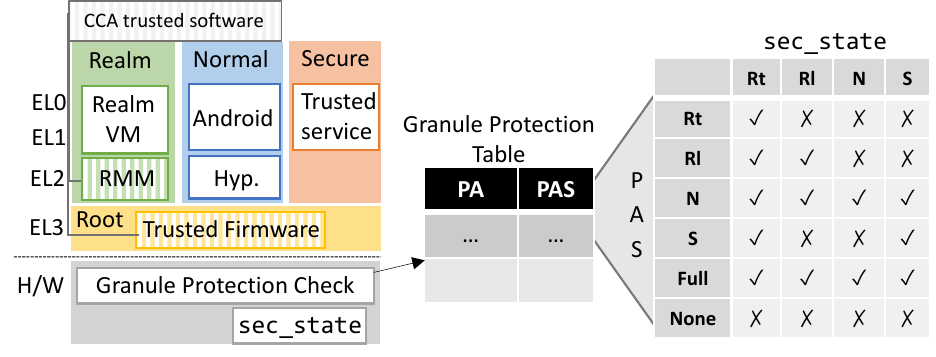} 
         \caption{Arm CCA architecture with CCA's trusted software with shaded background. Each core has a Granule Protection Check (GPC) which is programmed with a Granule Protection Table (GPT). Each core also has a \ssecstate register which indicates which world the core executes in : Root (Rt), Realm (Rl), Normal (N), or Secure (S). The GPT maps physical address to one of the six Physical Address Space (PAS) values. For each memory access, the GPC looks up the GPT and compares the PAS of the physical address to the value in the \ssecstate register.}
    \label{fig:cca}
 \end{figure}

\subsection{Background: Arm CCA}
\label{ssec:background-cca}
Arm Confidential Computing Architecture (CCA) extends Arm ISA with Realm Management Extensions (RME).
\noindent{\bf Worlds in Arm CCA.}
It introduces $2$ new worlds---realm and root (\cref{fig:cca}).   
With RME enabled, computation can execute in either normal, realm, root, or secure worlds. 
\cref{fig:cca} shows CCA's physical address space (PAS) access control based on the world of the software. 
Notably, all computation in the realm world is inaccessible to the normal world. 
The realm and secure worlds cannot access each other and the root world is inaccessible to any other worlds. 
Even in CCA the realm and secure worlds can access the normal world, thus 
inheriting TrustZone's overprivilege problems. 

\noindent{\bf Granules.}
CCA uses new hardware components called  Granule Protection Checks (GPCs) on each core, where a granule is the smallest block of memory that can be described (e.g., 4KB). 
The GPCs filter memory accesses from cores by looking up Granule Protection Tables (GPTs) which map granules in the physical address space to one of the $4$ worlds. 
In addition to the four worlds, the GPT can also indicate whether a physical address is fully accessible or not accessible to all software (see~\cref{fig:cca}).
The trusted firmware (\tf) executing in the root world programs the GPTs and ensures world isolation. 

\noindent{\bf Realm Management Monitor (RMM).}
CCA enables the creation of VMs in the realm world. 
The realm VMs are mutually distrusting and are isolated using the Realm Management Monitor (RMM).
The RMM is trusted software that executes in the realm world at EL2 and ensures realm VM isolation using stage-2 translation tables (S2 tables). 
To perform this isolation correctly, CCA specification defines invariants for the RMM.
Importantly, one of the RMM's invariants ensures that there can be no overlapping mappings in the realm world i.e., one realm world address is only mapped to one realm VM or the RMM. 
The RMM exposes an interface called the Realm Management Interface (RMI) to the hypervisor to create and manage realm VMs and a Realm Service Interface (RSI) to the realm VMs.
RMM and hypervisor invoke the \tf via a Secure Monitor Call (SMC).

\noindent{\bf Realm VM Creation and Attestation.}
To setup memory for a realm VM, the hypervisor first \emph{delegates} memory to the realm world by using an RMI call. 
On receiving this request, the RMM invokes the \tf to change the GPT world mapping for that physical address. 
Then, the hypervisor invokes RMI calls to add the delegated pages to the realm VM.
For this, the RMM updates the S2 tables according to its invariants to allow the realm VM access to the memory.
While adding the memory to the realm VM, the RMM also invokes the \tf to measure the pages using CCA hardware and adds the measurements to the realm VM's attestation report. 
Once all memory is added to the realm VM, the hypervisor finishes realm VM creation and boots it. 
After the realm VM boots, a remote verifier can request an attestation report from the realm VM. 
The realm VM invokes the RMM using RSI calls to get the report and sends it to the remote verifier.

\begin{figure*}
    \centering
     \includegraphics[scale=0.7]{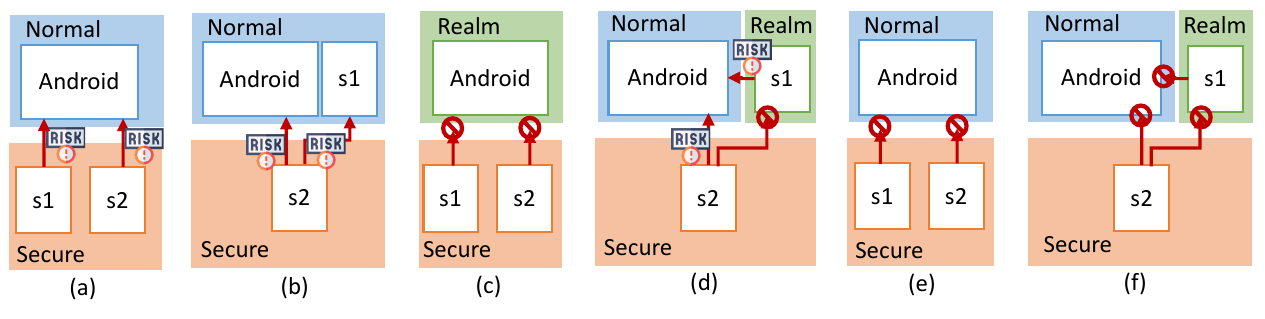} 
         \caption{Isolation in different settings where S1 and S2 are two \ta{s} such that: (a) \android executes in the normal-world and S1, S2 in the secure world. (b) \android and S1 are mutually isolated in the normal world. S2 is still overprivileged and can attack \android and S1. (c) Solution 1: move \android to realm world. (d) Solution 2: execute \sservices in the normal world (e) Solution 3: isolate \sservices in the secure world from \android using two-GPTs (f) \codename: executes \sservices in the realm world and ensures mutual isolation between the worlds.}
    \label{fig:settings}
 \end{figure*}

\noindent{\bf Realm VM Destruction.}
To destroy a realm VM, the hypervisor first stops the realm VM, destroys the S2 table entries, and then undelegates all realm memory by using RMI calls. 
To undelegate memory, the RMM invokes the \tf to update the GPT. 
According to CCA specifications, all realm memory is scrubbed before it is undelegated from the realm world. 
Once all realm VM memory is undelegated, the realm VM is effectively destroyed.

\noindent{\bf Runtime Memory Isolation.}
The GPCs filter every memory accesses from cores during runtime. 
To provide the GPC with the world of the core accessing the memory, the \tf programs a per-core register (\ssecstate) with the correct world on context switches (\cref{fig:cca}). 
To filter the memory accesses, the GPCs use this register and match it against the world of the physical address from the GPT. 
To speed up accesses to the GPT, every GPC maintains a GPC TLB where it caches GPT mappings. 
So, when the trusted monitor updates the GPT, it executes an instruction to flush all GPC TLBs as stated in the CCA specifications.

\subsection{Threat Model}
We assume a threat model where all software that executes in the normal world including \android and normal world hypervisor is untrusted. 
Similarly, we consider all software that executes in the secure world as untrusted. 
We deem the Arm CCA hardware as trusted and CCA's trusted software (RMM and \tf) to be implemented according to the specifications. 
Further, we assume mutual distrust between the \sservices. 
If an attacker compromises a sandbox then we cannot offer any more protections for it. 
Instead, we ensure that the compromise is limited to the sandbox and cannot affect other sandboxes or trusted software.
Protecting against side-channels and microarchitectural attacks is orthogonal to this work and we consider it out of scope.

\subsection{Potential Solutions}
\label{ssec:potential-solutions}

We explore different solutions using  Arm CCA to achieve the security primitives and highlight the challenges.

\subsubsection{Solution 1: Execute \android in Realm World} Arm CCA introduces a realm world where the normal world hypervisor can natively deploy confidential VMs (\cref{fig:cca}).
Further, any computation in the realm world is inaccessible to the normal and secure world.
We can leverage this new architecture to de-privilege the \ta{s}. 
At first glance, executing \android in the realm world while the \ta{s} execute in the secure world (\cref{fig:settings}(c)) seems like an obvious solution. 
With CCA, the secure world cannot access the realm world and vice versa. So, this design would ensure the mutual isolation primitive~\ref*{s:iso}.
Next, CCA specification delegates all resource management and scheduling tasks to a normal world hypervisor. 
Currently, the hypervisor is implemented in the \android kernel and is tightly coupled with it. 
So, to execute \android in the realm world, we would need to decouple the hypervisor from the \android kernel to execute separately in the normal world. 
Now, if we execute a part of \android (the hypervisor) in the normal world, it would no longer be fully isolated from the secure world breaking~\ref*{s:iso}. 

This design still executes \ta{s} in the secure world (\cref{fig:settings}(c)) where we should sandbox them to form \sservices and ensure~\ref*{s:sand}.
Several previous works have proposed solutions that sandbox the \ta{s} in the secure world (\cref{fig:secure-world}) which we can use to ensure~\ref*{s:sand}. 
However, because these solutions do not consider the problem  of secure interfaces we will need to carefully consider the interfaces and harden them to ensure~\ref*{s:int}.
Arm has announced support for TrustZone's secure world using CCA hardware~\cite{dynamic-trustzone}. 
While CCA supports hardware-based attestation measurements, it only defines these primitives for realm VMs and there is no specification for secure world attestation.
Therefore, to ensure~\ref*{s:att} for this design, we need to define new CCA interfaces to support hardware-based attestation primitives for \sservices in the secure world. 
We conclude that the changes we need to make to \android, CCA specification, and \ta{s} are invasive. And even with these drawbacks, it still cannot guarantee our desired security primitives.

\subsubsection{Solution 2: Execute \SServices in Realm World} 
CCA defines primitives for the realm VMs that ensure sandboxing and attestation. 
Thus, we now consider a solution that executes the \sservices in the realm world while \android executes in the normal world (\cref{fig:settings}(d)). 
This design achieves the sandboxing and attestation security primitives by executing the services in realm VMs. 
Furthermore, it standardizes the interface for realm VM and normal world communication which we can consider to ensure~\ref*{s:int}. 

However, this approach does not guarantee mutual isolation between the \sservices and \android. 
Architecturally, CCA allows the realm world to access all normal world memory. 
Therefore, simply moving all \sservices to the realm world will inherit the secure world's problem stemming from being over privileged \ta{s}.
Concretely, a normal world attacker can still compromise the new \sservices in the realm world and use the attacks from~\cref{ssec:tz-attacks}. 
Further, with this design the secure world continues to execute legacy \ta{s} which an attacker in the normal world can compromise. 
Therefore, this design does not mitigate the attacks that stem from the lack of least privilege notion in the realm or secure worlds. 
Despite this limitation, this solution is still a good first step towards ensuring our security primitives using Arm CCA. 
\subsubsection{Solution 3: \SServices in Secure World Isolated from Normal World}
The main problem with Solution 2 is that it does not solve the problem of privileged \sservices in the secure and realm worlds. 
Although CCA isolates realm and secure worlds, isolating normal world from malicious privileged services (in the realm and secure worlds) to ensure~\ref*{s:iso} requires further investigation. 

\noindent{\bf Use CCA to De-privilege Realm and Secure World.}
To ensure mutual isolation, we need to create two different spatial views of memory: one that allows access to normal world memory (for \android), and one that blocks accesses to normal world memory (for \sservices in realm or secure worlds).
By default, CCA uses just one view of memory for all cores by programming them with the same GPT. 
To create our different views of memory, we can create 2 different GPTs that allow and disallow access to normal world memory. 
Then, we program the cores that execute in the normal world (\android) with the GPT that allows normal world memory access. 
On the other hand, we program cores that execute in the realm or secure world (\sservices) with the GPT that blocks normal world memory access. 
As a consequence, if software executing in realm or secure worlds originate accesses to normal world memory, the CPU raises a Granule Protection Fault (GPF) and the access is stopped. 
With this mechanism, we enforce the notion of least privilege by limiting the memory that \sservices can access in the normal world, fulfilling~\ref*{s:iso}. 

\noindent{\bf Solution Description.}
With this $2$ GPT setup, consider a design where the \sservices execute in the secure world and \android executes in the normal world (\cref{fig:settings}(e)). 
The $2$ GPT setup ensures that \android is isolated from the secure world for~\ref*{s:iso}. 
However, we still need to ensure the other security primitives for sandboxing, secure interfaces, and attestation. 
Like Solution 1, we can use previous works to ensure sandboxing in the secure world. 
But securing interfaces and hardware based attestation will require overhauling changes to the CCA specification and the \sservices. 
On the other hand, the realm world with its native support for isolated VM execution, standardized interfaces, and attestation primitives is therefore a more amenable choice to execute \sservices.

%% file: sections/04_overview.tex
\section{\codename Overview}
Following the description of the Arm CCA and analysis of potential solutions, here we present \codename, that uses Arm CCA to enable \sservice execution with \android on mobile platforms. 
For this, \codename executes \android in the normal world and \sservices in the realm world (\cref{fig:settings}(f)) and uses 2 GPTs to ensure mutual isolation between the normal, realm, and secure worlds. 
\begin{figure*}[t]
    \centering
     \includegraphics[scale=0.65]{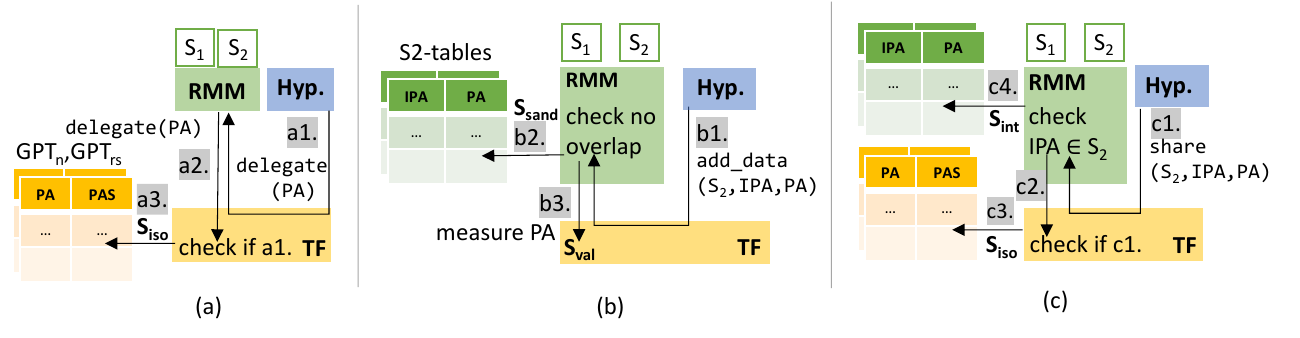} 
         \caption{Enforcing \codename primitives during \sservice creation. S1 and S2 are \sservices, PA is physical address, IPA is intermediate physical address (a) a1. Hypervisor delegates PA to realm; a2. RMM calls TF to delegate PA to realm; a3. TF checks and then updates GPT to ensure~\ref*{s:iso}. (b) b1. Hypervisor adds data to \sservice S2; b2. RMM checks that the PA is not assigned to any other \sservice for~\ref*{s:sand}, then updates S2-tables; b3. then calls the TF to measure the granule~\ref*{s:att}. (c) c1. Hypervisor calls create shared memory for \sservice; c2. RMM checks and invokes the TF to update GPT; c3. TF checks then updates GPT for~\ref*{s:iso} and then returns to the RMM; c4: RMM updates the S2-tables to enforce \ref*{s:int}.}
    \label{fig:design}
 \end{figure*}
Then, we develop an ecosystem where \android apps in the normal world create and communicate with \sservices while ensuring our security primitives. 
With \codename, an \sservice can execute a bare-metal binary, a binary that runs on top of a thin library OS, or a full VM that executes a Linux kernel with userspace applications. 
In all these cases, the \sservice should contain a component that can perform RSI calls to invoke the RMM. 
To create \sservices, an \android app invokes \android APIs and sends the code and data to execute in the \sservice along with a manifest. 
The manifest contains details of \sservice memory regions, and location and size of shared memory used to create \sservice-\android interfaces.
\android invokes the normal world hypervisor with these details to allocate memory and resource and spawn the \sservice.

\codename leverages the Arm CCA platform to deploy the \sservices. 
With CCA, the RMM exposes RMIs to create and manage VMs in the realm world. 
In \codename, \android's normal world hypervisor invokes these RMIs to create \sservices in the realm world. 
To ensure our security primitives, \codename modifies Arm CCA's VM creation process to: (a) setup the $2$ GPTs to ensure mutual isolation, (b) perform local attestation of the \sservice and assist with remote attestation using CCA's hardware, and (c) create secured shared memory that \android can use to communicate with the \sservice. 
To destroy an \sservice, \codename uses CCA's realm destruction flow.
This ensures that all pages that belong to the realm VM are undelegated, scrubbed and transferred back to the normal world. 
\cref{tab:changed-interfaces} shows an overview of all the interfaces and CCA components that \codename changes.

\subsection{Mutual Isolation and Sandboxing}
\label{ssec:mutual-iso-sandboxing}

\begin{table}[]
\caption{Interfaces that \codename adds or changes in CCA. Where N: New and E: Existing.}
\label{tab:changed-interfaces}
\resizebox{\columnwidth}{!}{%
\begin{tabular}{@{}llll@{}}
\toprule
{\bf Action}                                                                                                                   & {\bf Interface}                                                                                                                                                                                                                  & {\bf Components}                                              & N/E                                              \\ \midrule
\textbf{\em Memory delegation:} update 2 GPTs                                                                                & \begin{tabular}[c]{@{}l@{}}$\tt{rmi\_granule\_delegate}$, \\ $\tt{rmi\_granule\_undelegate}$\end{tabular}                                                                                                              & TF                                                      & E                                                \\ \midrule
\begin{tabular}[c]{@{}l@{}}\textbf{\em Memory sharing:} update 2 GPTs\\ \\ invoke $\tt{smc\_2gpt\_ns\_share}$\end{tabular} & \begin{tabular}[c]{@{}l@{}}$\tt{smc\_2gpt\_ns\_share}$\\ $\tt{rmi\_map\_unprotected}$, $\tt{rmi\_data\_create}$, \\ $\tt{rmi\_realm\_create}$, $\tt{rmi\_rec\_create}$, \\ $\tt{rmi\_rec\_enter}$\end{tabular} & \begin{tabular}[c]{@{}l@{}}TF\\ \\ RMM, TF\end{tabular} & \begin{tabular}[c]{@{}l@{}}N\\ \\ E\end{tabular} \\ \midrule
\begin{tabular}[c]{@{}l@{}}\textbf{\em Exclusive access:} update \gptns\\ call $\tt{smc\_2gpt\_ex\_access}$\end{tabular}   & \begin{tabular}[c]{@{}l@{}}$\tt{smc\_2gpt\_ex\_access}$\\ \rsiexaccess\end{tabular}                                                                                                                                      & \begin{tabular}[c]{@{}l@{}}TF\\ RMM\end{tabular}        & \begin{tabular}[c]{@{}l@{}}N\\ N\end{tabular}    \\ \midrule
\textbf{\em MMIO protection:} mark MMIO                                                                                      & \rsimmio                                                                                                                                                                                                                   & RMM                                                     & N                                                \\ \bottomrule
\end{tabular}%
}
\end{table}

\begin{figure}[t]
    \centering
     \includegraphics[scale=0.57]{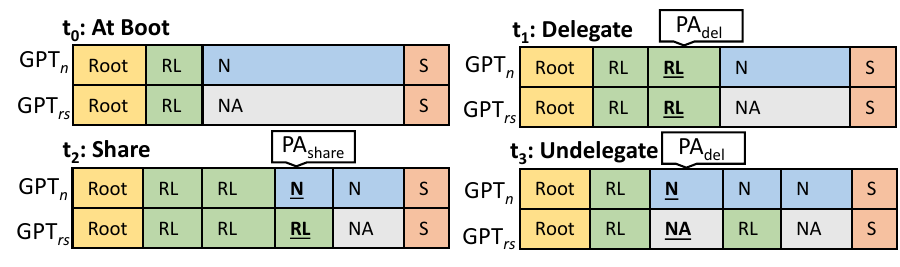} 
         \caption{Changes to the 2 GPTs by the \tf from time t$_0$ to t$_3$. Root (in yellow), Realm (RL in green), Normal (N in blue), Not-accessible (NA in grey), and Secure (S in orange). \textbf{t$_0$:} at boot \tf marks the normal world memory as not-accessible in \gptrs. \textbf{t$_1$:} to delegate granule PA$_{del}$ the \tf marks it as realm in both GPTs. \textbf{t$_2$:} to share granule PA$_{share}$ the \tf marks it as normal in \gptns and realm in \gptrs. \textbf{t$_3$:} to undelegate PA$_{del}$ the \tf marks it normal in \gptns and not-accessible in \gptrs.}
    \label{fig:gpt}
 \end{figure}

\codename ensures~\ref*{s:sand} for the \sservices by using the RMM's realm VM memory invariants. 
The RMM guarantees that two realm VMs do not have overlapping memory in the realm world. 
\codename uses this guarantee to ensure~\ref*{s:sand} for \sservices (\cref{fig:design}(b)). 
With this, any bugs in an \sservice are contained to its sandbox and cannot escape to compromise other \sservices. 
Next, to ensure mutual isolation between the \sservices in the realm world and \android in the normal world, \codename uses the $2$ GPT setup. 
To create the two spatial memory views \codename uses 2 GPTs; \gptns for the normal world, and \gptrs for the realm and secure worlds. 
The key difference between these GPTs is that \gptrs marks all normal world memory as not-accessible (\cref{fig:gpt}(a)). 
During runtime, \codename uses \gptns for all normal world cores that execute \android and the hypervisor. 
Similarly, it uses \gptrs for all cores that execute realm and secure world software.

\noindent{\bf Creating $2$ GPTs.} 
By default, the \tf creates a GPT during boot that is used for all cores. 
For this, it allocates space in root memory and configures the GPT to designate memory in the realm world for the RMM, in the normal world for \android, and in the secure world for any secure world hypervisors, OSes, and \ta{s}.
\codename reuses this original GPT in CCA as \gptns. 
To create \gptrs, \codename extends this boot process in the \tf and allocates additional space in root memory for \gptrs. 
Then, \codename marks all normal world memory in this GPT as not-accessible (see~\cref{fig:gpt}).

\noindent{\bf Managing $2$ GPTs for \SServices.} 
With CCA, when the hypervisor delegates memory to the realm the RMM invokes the \tf to update the GPT to designate the granule to realm (\cref{fig:design}(a)).
Similarly, when the hypervisor undelegates realm memory to the normal world, the RMM invokes the \tf to update the GPT. 
\codename extends this granule delegation and undelegation flow to maintain the two GPTs for \sservices.
During \sservice creation, when the RMM invokes the \tf to delegate a granule to realm, the \tf marks the granule as realm in both \gptns and \gptrs. 
Once the \sservice boots, it can request more memory from the hypervisor through the RMM by using RSI calls.
In response, the hypervisor performs RMI calls to delegate memory to the realm before assigning it to the \sservice. 
As before, for the realm memory delegate operation, the \tf marks the granule as realm in both the GPTs. 

If the \sservice relinquishes realm memory or is destroyed, the RMM invokes the \tf to undelegate the realm memory to the normal world. 
For this operation, \codename changes the \tf to mark the granule as normal world in \gptns and not-accessible in \gptrs (see~\cref{fig:gpt}). 
CCA specification states that realm memory is always cleaned before it is undelegated to the normal world. 
Therefore, \codename  
does not need to perform any additional operations when relinquishing \sservice memory ensuring that it does not leak.

To ensure that both the realm and normal worlds have consistent memory views and guarantee hardware-enforced mutual isolation, the \tf checks that both the RMM in the realm world and the hypervisor in the normal world agree to the memory delegate or undelegate operation. 
In CCA, the hypervisor always originates the delegate or undelegate requests by using RMI calls that are routed through the \tf (\cref{fig:design}(a)). 
To ensure consistent views, \codename changes the \tf to store these RMI parameters before routing them to the RMM. 
Then, when the RMM invokes the \tf to delegate or undelegate memory, the \tf first checks that the hypervisor had requested it, and only then performs the operation. 

\subsection{Interfaces}
Our security primitives which ensure mutual isolation and sandboxing, provide \codename with a clean abstraction to investigate interface security. 
Because the \sservices are mutually isolated from \android, they cannot arbitrarily access normal world \android apps. 
Instead, \codename forces them to use a shared memory region which can only be created using the RMM.
This gives \codename an opportunity to use the RMM to mediate such shared memory setup for communication interfaces between \sservices and \android applications enforcing~\ref*{s:int}. 
With \codename, \sservices use the shared memory interface to either communicate with \android apps using a standard RPC interface or use the normal world hypervisor's virtual device support (e.g., virtio for block devices and network).
Next, we explain how we build \codename's shared memory based interfaces. 
For this, we draw insights from several prior works in non-CCA setting~\cite{avf, optee, keystone, elasticlave}.

\subsubsection{Communications with \android Apps}
\label{sssec:communication-with-android-apps}

We build a secure shared memory region that can be used by \sservices to communicate with \android. 

\noindent{\bf Setting up Shared Memory.}
The \android app that requests \sservice creation specifies the size of the shared memory region. 
The \sservice uses this region to allocate all objects for communication with \android. 
\codename sets up this fixed shared memory region with the right parameters during \sservice creation.
This requirement for a fixed shared memory region is stricter than the current CCA specification.
Currently, CCA allows realm VMs to dynamically decide which pages in the realm VM memory are shared with the normal world. 
\codename strengthens this by requiring that all shared memory be set up in a contiguous space during \sservice creation. 
This hardens the interface by ensuring that \sservice does not unintentionally expose its data to the normal world (e.g., exposing a full page as shared memory to the hypervisor while only intending to share a small object in the page). 

To create the shared memory region, the hypervisor invokes an RMI call with the information of the memory pages to share. 
Before forwarding this RMI to the RMM, the \tf logs this request in its root memory. 
Then, to create the shared memory region, the RMM invokes the \tf to mark the region as normal world in \gptns and realm world in \gptrs (see~\cref{fig:gpt}). 
Before performing this operation, the \tf first checks that the hypervisor had invoked the RMI to share the memory pages (\cref{fig:design}(c)). 

\noindent{\bf Protecting Shared Memory.}
\codename hardens the interface by using the RMM to mark the shared memory region as non-executable and enable mechanisms for exclusive access to the \sservice. 
To ensure that the shared memory region is non-executable, \codename extends the RMM's shared memory creation mechanisms and marks the region as non-executable in the \sservice's S2 tables.

An \sservice can perform \sservice-specific interface checks on the data that it reads or writes to the shared memory region. 
Naively performing these checks can leave the \sservice vulnerable to TOCTOU attacks. 
For example, an \sservice checks an object in shared memory for bad values (e.g., bounds checks). Once the check passes, before the \sservice can use the value, the hypervisor changes it mounting a TOCTOU attack. 
Therefore, exclusive access to shared memory is important while performing interface checks to prevent such TOCTOU attacks. 
In our setting, we cannot enable this by default using the RMM, because these interface checks and when to perform them are linked to the semantics of the \sservices and the interfaces they use. 
Instead, \codename enables a mechanism that the \sservice can use to lock down shared memory pages for exclusive access. 
To implement this exclusive access mechanism, \codename introduces a new RSI call. 
The \sservice can invoke this RSI with information on the pages of shared memory to enable exclusive access on.
On receiving this request, the RMM uses the \tf to mark the requested regions of memory as not-accessible in \gptns, preventing the normal world software from accessing it. 
The \sservice can disable the exclusive access by invoking the RSI again which marks the regions back to normal world in \gptns using the \tf.

\noindent{\bf Standard Interfaces for Communication.}
Just a shared memory region is insufficient to enable expressive communication between \android and the \sservices. 
Several previous works have proposed protocols that work over shared memory (e.g., \android's Binder RPC for communication between different processes, Native Client's RPC). 
\codename's shared memory abstraction is general enough and \sservices can use these standard protocols to communicate over it enforcing~\ref*{s:int}. 
In our implementation, we pick one concrete instance of \android's Binder RPC to show \codename's expressiveness. 
To use the Binder, the \android app and the \sservice agree on an interface definition (types of commands, data types of objects sent/received). 
We observe that this definition is rich enough to automatically invoke the exclusive access RSI calls when data is sent over the Binder interface. 
Therefore, we implement these calls as part of the \sservice in our implementation. 
Furthermore, this interface has been extensively used in \android app development for inter-process communication and for VM communication~\cite{feng2016bindercracker}. 
We show that \codename's design is compatible with this standard interface and strongly recommend that \sservices use it. 
\Sservices that use this interface can benefit from its mature testing infrastructure (e.g., fuzzers, analysers) further strengthening \sservice security and enforcing~\ref*{s:int}.

\subsubsection{Virtio Device Support}
\label{ssec:virtio}
\codename supports attaching selected untrusted virtio devices, which are managed by the hypervisor, to \sservices. 
To use virtio devices, \sservices need to send data to the devices through the untrusted hypervisor. 
For some devices (e.g., keyboard, display) without cryptographic endpoints, \sservices will need to send this data in plaintext which will open them up to attacks from the hypervisor. 
In \codename, we observe that \sservices mainly use virtio for networking and storage in block devices (see~\cref{sec:case-studies}). 
These virtio devices can be secured using network encryption and encrypted storage. 
For these virtio devices the hypervisor merely acts as a relaying entity. 
Even if the hypervisor changes their data in any way, encryption and integrity protection will detect such tampering. 
Therefore, in \codename we use local attestation to ensure that, if any, only virtio devices for networking and block storage can be attached to an \sservice (see~\cref{ssec:attestation}).

\noindent{\bf Allocating Virtio Device Queues.}
The untrusted hypervisor manages virtio devices for the \sservice. 
For this, the hypervisor sets up queues (virtqueues) in the shared memory region that \codename creates. 
The \sservice writes into these queues to send data to the hypervisor to perform device operations (e.g., send data to the device with DMA). 
Typically, virtio implementations assume that the hypervisor can directly access VM memory. 
So, the VM writes pointers to objects in its memory in the virtqueues for the hypervisor to dereference. 
To allow this functionality, in \codename we should mark the memory that holds these objects as shared with the hypervisor. 
We see that these objects might not be page aligned while \codename shares memory with the hypervisor at page granularity. 
Therefore, by sharing the object's memory we might leak \sservice memory to the hypervisor.  
To prevent this and enforce \ref*{s:int}, \codename fixes the shared memory region during \sservice creation. 
Then, all objects that might have pointers in the virtqueues should be allocated in this fixed shared memory region. 

\noindent{\bf MMIO protection.}
MMIO reads and writes to the virtio devices are performed by the hypervisor. 
For an MMIO write, the hypervisor needs to write the register value to the device. 
Similarly, for an MMIO read, the hypervisor should be able to copy the value from the device into the \sservice's register. 
With CCA, the hypervisor cannot directly access the realm memory to perform these operations. 
Instead, to provide this functionality, when the \sservice accesses memory-mapped regions (for MMIO read/write) of the virtio devices, the hardware generates a data abort that the RMM traps on. 
To perform the MMIO operations, the RMM sends data about the faulting instruction (e.g., register values for MMIO write) to the hypervisor. 
Furthermore, for MMIO read, the RMM copies data from the hypervisor into the \sservice register state.  
Previous works have shown that such interfaces can be abused if they are not correctly implemented~\cite{wesee-oakland}.
Currently, the RMM checks that the realm VM exited because of a data abort (i.e., because of a read or write to memory) and only then processes the MMIO operation. 
\codename strengthens this and ensures~\ref*{s:int} by requiring the \sservice to explicitly mark regions of its memory as MMIO regions using a new interface in the RMM.
With this, we extend the RMM to check if the address that caused the data abort was marked as an MMIO region by the \sservice and only then perform the MMIO data transfers.

\subsubsection{\bf Interrupts}
Ahoi attacks abuse the notification interface (e.g., interrupts, exceptions) between trusted and untrusted software by triggering expressive handlers~\cite{heckler-usenix, wesee-oakland}. 
With CCA, the interrupts for the realm VMs are managed by the hypervisor.
Therefore, the hypervisor can inject interrupts into the \sservice at any point during their execution 
On Arm, exceptions (e.g., divide-by-zero, data abort) are injected by trusted hardware and cannot be faked by the hypervisor. 
Other interrupts like timer  do not typically have expressive handlers for an attacker to exploit. 
All that remains is device originated interrupts,  
which the RMM filters by default and hence cannot be abused by attackers to mount Ahoi attacks~\cite{heckler-usenix}. 
\codename does need two virtio devices (block storage and network) that are attached to \sservices. 
The attacker can inject interrupts for these devices, but since \codename uses encryption and integrity protection, the victim \sservice will detect it (e.g., failed decryption and integrity checks). 

In summary, \codename uses the RMM to build secure shared memory and enforces security measures including fixed shared memory regions, non-executable data, selected virtio interfaces, and MMIO protection. 
Besides these, \codename enables \sservices to turn on/off exclusive access on shared memory which allows them to perform \sservice-level semantic checks when sending/receiving data from the untrusted hypervisor.

\subsection{Attestation and Validation}
\label{ssec:attestation}
CCA provides hardware primitives to perform local and remote attestation for realm VMs. 
In \codename, we use these hardware primitives to enforce~\ref*{s:att}.

\noindent{\bf Local Attestation.}
In CCA, the hypervisor can use RMI calls to launch and execute arbitrary VMs. 
On a mobile platform, this might not be desirable as the software is usually checked and validated by \android.
Furthermore, for \sservice security, we need mechanisms to ensure that only selected virtio devices (\cref{ssec:virtio}) are attached to the \sservice. 
\sservice validation can be performed by \android before they are launched, but in our setting, \android is untrusted. 
Instead, \codename uses the RMM to perform local attestation and enforce validation rules for virtio devices.

\noindent{\bf Remote Attestation.}
As explained in~\cref{ssec:background-cca}, CCA hardware populates attestation reports for the \sservices when they are created. 
This report includes measurements of all \sservice memory and platform software (RMM and \tf). 
A remote verifier can query this report (see~\cref{ssec:background-cca}) and ensure that the \sservice is booted correctly on a trustworthy platform ensuring~\ref*{s:att}.

%% file: sections/06_security-analysis.tex
\section{Security Analysis}

\noindent{\bf Attacks from Secure World.}
Compromised \ta{s} in the secure world can try to access \sservices in the realm world. 
But, CCA's GPC checks prevent the secure world from accessing realm world memory. 
The compromised \ta{s} can try to access shared memory that \sservices setup with the normal world but CCA hardware stops this as this memory is marked as realm world in \gptrs.

\noindent{\bf Attacks from \android and the Hypervisor.}
Normal world software (\android apps, VMM, or the hypervisor) can try to launch \sservices that do not conform to the platform's validation rules. 
Such attempts will be detected and discarded by \codename during \sservice boot (\ref*{s:att}). 
Malicious \android apps can try to launch attacker controlled \sservices to compromise other \sservices. 
This attack is stopped by \codename's sandboxing primitive (\ref*{s:sand}). 
They can also try to attack the \android kernel in the normal world which is stopped by the mutual isolation primitive (\ref*{s:iso}) that \codename guarantees. 
When an \android app requests \sservice creation, the hypervisor can create \sservices with incorrect configurations, devices, or malicious code or data. 
However, these bad startup settings will be detected (\ref*{s:att}) either during the local attestation process or during remote attestation.
The hypervisor can try to mount a split-view attack where cores see different views of the GPTs by not synchronizing all cores or trying to update the GPT simultaneously from different cores. 
In CCA, the updates to the GPT are synchronous i.e., the RMM waits for the \tf to return after it makes the request. 
Before updating the GPT, the \tf always acquires a spinlock ensuring that only one core can update the GPT at any point in time. 
Then before returning to the RMM, the \tf always executes an instruction that flushes the GPC TLBs on all cores forcing them to refetch the updated GPT entries. 

\noindent{\bf Interface Attacks.}
Normal world software can try to compromise \sservices by using \codename's shared memory interfaces or the virtio device queues. 
\codename stops attacks where the attacker injects code into the shared memory region and executes it in the \sservice by ensuring that the shared memory regions are always non-executable. 
Further, \codename stops attacks from normal world software that abuses virtqueues implementations to de-reference arbitrary \sservice memory by fixing the shared memory region during \sservice creation. 
Besides these global mechanisms that \codename employs, we cannot control  what data is sent over the shared memory channels and how they are used by the \sservices or the normal world. 
To harden the interface further, \codename enables exclusive access to the shared memory regions that the \sservices can turn on or off. 
This allows \sservices to deploy context-specific interface hardening rules. 
Attackers can compromise buggy \sservices that do not employ security-relevant features (e.g., exclusive access to shared memory, stack canaries, ASLR). 
\codename cannot completely stop attackers from compromising these buggy \sservices even if it limits the interfaces available to the attackers. 
But, \codename's sandboxing and mutual isolation primitives  ensure that these attackers cannot compromise other \sservices (\ref*{s:sand}) or the normal world software (\ref*{s:iso}).

\noindent{\bf Compromised \SServices.}
They can try to bypass the \android permission models to get access to data (e.g., contacts) or devices (e.g., camera) in the normal world. 
However, because \codename ensures mutual isolation, they cannot directly access \android memory to bypass permission checks.
Compromised \sservices can try to use their interface with the \android app to bypass the permission checks, which \android stops as it imposes the checks on all \android apps. 
Compromised \sservices can try to attack other \sservices but are stopped by \codename's sandboxing primitive. 
They can also try to escalate privilege to the RMM using the RMM's interface. 
However, with CCA the RMM's interface with the \sservice is small and strictly regulated, always checks the inputs, copies only fixed data from the realm VMs, and is deemed secure.
In \codename, we ensure that the security of the interface by following CCA's security practices. 
Compromised \sservices can try to attack \android in the normal world which is stopped by~\ref*{s:iso}, by marking Android memory as inaccessible.

\noindent{\bf Malicious Devices.}
The hypervisor can corrupt or tamper the virtio device data. 
\codename only allows block devices and network devices as virtio devices which we are secured using disk and network encryption, thus stopping this attack. 
Other integrated devices can try to access \sservice memory, but are stopped by CCA because the SMMU's GPC is programmed with \gptns by default, which does not allow these devices access to realm memory. 

\noindent{\bf Physical Attackers.} TrustZone, and by extension all prior works, does not protect against a physical attacker that can snoop on the DRAM, mainly due to lack of memory encryption. \codename can use Arm CCA Memory Encryption Contexts (MEC) to generate per-\sservice identities (MEC IDs)~\cite{arm-mec}. So the hardware  uses a unique key for each \sservice to encrypt and integrity protects the data before DRAM.

%% file: sections/07_implementation.tex
\section{Implementation}
To implement \codename we pick Linaro's CCA stack~\cite{linaro-cca} with implementations for the RMM (v1.0-eac5)~\cite{rmm-linaro}, \tf (v$2.10$)~\cite{tf-linaro}, and a Linux kernel (v$6.7$-rc4) with patches for CCA~\cite{linux-linaro}. We use Android Open Source Project (AOSP) v$13.0.0$\_r$12$~\cite{aosp-13} with Android's patched Linux kernel (Common kernel v15-6.6)~\cite{aosp-kernel-15}.\footnote{We refer to this patched Linux kernel as Android kernel.}

\subsection{\SServices in the Realm World}
\codename needs to execute a user-space binary and a kernel as part of the \sservice. 
We consider 
existing \android services that can benefit from \codename's protections. 
In particular, \android enables running protected VMs (pVMs) in the normal world using a trusted hypervisor (called pKVM) as part of the new \android Virtualization Framework (AVF). 
While a pVM can run any OS and application in the VM, the \android platform provides a minimal runtime which is a stripped down version of \android called Microdroid. 
We take existing apps that run Microdroid VMs and port them to run as \codename's \sservices. 
In the VM, \android executes Microdroid in EL0, an Android kernel in EL1, and uses a modified uboot as the VM's bootloader.
To build, deploy, and manage the Microdroid VM, the \android platform uses a Virtual Machine Monitor called crosvm. 
We modify the \android Linux kernel, the bootloader, and crosvm to launch Microdroid VMs as \sservices in the realm world.

\noindent{\bf Adding CCA Support to the \android Kernel.} 
To build \codename we first need to enlighten \android's Linux kernel with CCA support. 
So we cherry-pick CCA support from Linaro's CCA Linux implementation into our \android kernel, which adds $2554$~\loc.
The patchset allows us to use the same kernel in the host (i.e., with \android) in the normal world and as a guest in the realm world. 
This enlightened \android kernel when executed in the realm world performs RSI calls to communicate with the hypervisor through the RMM. 
We verify that our changes to the guest kernel are compatible with Microdroid and boot a VM in the normal world. 
This VM executes Microdroid in EL0 with our modified \android kernel in EL1.
This setup also serves as a baseline to evaluate \codename against.

\begin{table*}[]
\caption{Changes to create and boot \sservices in realm world}
\label{tab:crosvm}
\centering
\resizebox{2\columnwidth}{!}{%
\begin{tabular}{@{}llll@{}}
\toprule
\bf{Operation}                                                                                           & \bf{Component}                                                    & \bf{RMI/RSI call}                                                           & \bf{Description}                                                                     \\ \midrule
\begin{tabular}[c]{@{}l@{}}Allocate memory for the VM and transfer it to realm world\end{tabular} & crosvm                                                       & $\tt{rmi\_granule\_delegate}$                                                 & \begin{tabular}[c]{@{}l@{}}Delegate a granule to the realm world\end{tabular} \\ \midrule
\begin{tabular}[c]{@{}l@{}}Deploy initial Data into Realm\end{tabular}                           & crosvm                                                       & $\tt{rmi\_data\_create}$                                                      & \begin{tabular}[c]{@{}l@{}}Copy Data from NS to Realm memory\end{tabular}     \\ \midrule
\begin{tabular}[c]{@{}l@{}}Create PA to Realm IPA mappings\end{tabular}                           & \begin{tabular}[c]{@{}l@{}}crosvm/hypervisor\end{tabular} & $\tt{rmi\_rtt\_create}$                                                       & \begin{tabular}[c]{@{}l@{}}Create Realm Translation table\end{tabular}      \\ \midrule
\begin{tabular}[c]{@{}l@{}}Establish VM identifier between hypervisor and RMM\end{tabular}      & crosvm                                                       & $\tt{rmi\_realm\_create}$                                                     & Creates a Realm                                                                 \\ \midrule
\begin{tabular}[c]{@{}l@{}}Create Storage for Realm vCPU data\end{tabular}                        & crosvm                                                       & $\tt{rmi\_rec\_create}$                                                       & \begin{tabular}[c]{@{}l@{}}Creates a Realm Execution Context\end{tabular}     \\ \midrule
Finalize creation of Realm                                                                          & crosvm                                                       & $\tt{rmi\_realm\_activate}$                                                   & Activates a Realm                                                               \\ \midrule
Execute a Realm                                                                                     & \begin{tabular}[c]{@{}l@{}}crosvm/hypervisor\end{tabular} & $\tt{rmi\_rec\_enter}$                                                        & Enters a REC                                                                    \\ \midrule
\begin{tabular}[c]{@{}l@{}}Share memory from the host with a Realm\end{tabular}                   & crosvm                                                       & \begin{tabular}[c]{@{}l@{}}$\tt{rmi\_rtt\_map\_unprotected}$\end{tabular} & \begin{tabular}[c]{@{}l@{}}Creates a non-protected IPA mapping\end{tabular} \\ \midrule
\begin{tabular}[c]{@{}l@{}}Set host memory access permissions for Realm memory\end{tabular}      & \begin{tabular}[c]{@{}l@{}}uboot/guest kernel\end{tabular} & $\tt{rmi\_rtt\_set\_ripas}$                                                   & \begin{tabular}[c]{@{}l@{}}Changing Realm IPA state\end{tabular}              \\ \midrule
\begin{tabular}[c]{@{}l@{}}Access device memory in unprotected IPA range\end{tabular}            & uboot                                                        & N/A                                                                    & N/A                                                                             \\ \midrule
\begin{tabular}[c]{@{}l@{}}Perform exclusive memory access of device memory\end{tabular}          & \begin{tabular}[c]{@{}l@{}}uboot/guest  kernel\end{tabular} & $\tt{rsi\_ex\_access}$                                                        & \begin{tabular}[c]{@{}l@{}}Turn exclusive page access on or off\end{tabular}  \\ \midrule
\begin{tabular}[c]{@{}l@{}}Limit MMIO emulation to specific memory regions\end{tabular}           & \begin{tabular}[c]{@{}l@{}}uboot/guest kernel\end{tabular} & $\tt{rsi\_mmio}$                                                              & \begin{tabular}[c]{@{}l@{}}Change memory affiliation\end{tabular}            \\ \bottomrule
\end{tabular}
}
\end{table*}

\noindent{\bf Launching \SServices with \android.}
For \codename's \sservices, we need to create Microdroid VMs in the realm world. 
To create the realm VMs, we modify crosvm with $469$\,\loc to trigger RMI calls.
With our changes crosvm invokes \android kernel's KVM to perform RMI calls during \sservice creation and management (see~\cref{tab:crosvm} for more details).
We extend \android's bootloader to correctly boot Microdroid in the realm VM.
For this, we patch the bootloader to perform RSI calls instead of the hypervisor calls that it performs (see~\cref{tab:crosvm}). 
In total, we change $543$~\loc in the bootloader. 
These changes allow us to boot a realm VM with our patched CCA Android kernel and Microdroid.

\subsection{Implementing \codename security primitives.}
We modify the RMM and \tf for our primitives (\cref{tab:changed-interfaces}).

\subsubsection{RMM}
We change 304~\loc in the RMM to implement the security primitives.

\noindent{\bf Creating and Maintaining Shared Memory Regions.}
In CCA, several RMI calls (see~\cref{tab:changed-interfaces}) create shared objects with the normal world during \sservice creation. 
In \codename's RMI handlers for these calls, we first check that these objects are in pages that are contiguous with any existing shared memory pages. 
Then, we invoke a newly introduced $\tt{smc\_2gpt\_ns\_share}$ call to the \tf to update the GPTs. 
When the \tf returns, we update the S2 table entries for the shared memory pages to be non-executable. 
Because \codename expects the hypervisor to setup a fixed shared memory region during boot, we disable this shared memory creation mechanism and the SMC invocation once the \sservice boots. 
To enable exclusive access to shared memory we introduce a new RSI call (\rsiexaccess) that takes as argument a list of guest physical addresses and a parameter indicating whether to turn exclusive access on/off for the pages. 
In the RSI handler, the RMM first checks that these pages belong to the \sservice that invoked the RSI and that they are shared pages. 
Then, to turn the exclusive access on/off on these pages, the RMM invokes a new $\tt{smc\_2gpt\_ex\_access}$ call to the \tf to update the GPTs (see~\cref{sssec:communication-with-android-apps}).

\noindent{\bf MMIO Protection.}
In \codename, the \sservice needs to explicitly mark pages for MMIO emulation. 
For this, we implement a new RSI call (\rsimmio) which takes as input the pages to mark for MMIO emulation. 
In response to the RSI, we change the RMM to store this information in the \sservice context. 
Then, before performing the MMIO emulation for an \sservice, the RMM first checks that the pages were marked for MMIO emulation using \rsimmio and only then copies data to or from the untrusted hypervisor.  

\subsubsection{\tf}
We change $610$~\loc in the \tf to implement \codename's security primitives. 

\noindent{\bf Creating and Maintaining GPTs.}
We create and populate the $2$ GPTs at  platform boot. 
For the new GPT, we allocate an additional $2$\,MB  memory in the root world.
Then, when the RMM uses the SMC to invoke the \tf to delegate, undelegate, or share realm memory, we have to first check that the hypervisor requested this operation and then update the GPTs. 
To check, we observe that the RMM only invokes these SMC calls in response to the hypervisor's RMI calls which are always routed through the \tf. 
So, we change the \tf to store these RMI parameters from the hypervisor before forwarding them to the RMM. 
Then, when the RMM invokes the \tf, we lookup the information to check that the hypervisor truly invoked the right RMI with the corresponding arguments (addresses to delegate, undelegate, or share). 
If the checks pass, we update the GPTs. 

We implement a new SMC in the \tf for the RMM to turn exclusive access on/off to shared memory pages (see~\cref{sssec:communication-with-android-apps}). 
Here, we do not need to synchronize views with the hypervisor as  the hypervisor has already acknowledged 
this memory is shared with the \sservice.
Instead, we check that the requested pages are in the shared state i.e., normal world in \gptns and realm world in \gptrs and then update the GPTs.
On context switches between the realm, normal, or secure world we change the \tf to program the $\tt{gptbr\_el3}$ register with the corresponding GPT. 
After changing the register, we ensure that the GPC TLBs are flushed according to CCA specifications.

\subsection{Local SBS Attestation}
\label{appx:tsfw}

To perform local attestation and to validate the \sservice, the RMM uses a trusted bootloader. 
\codename loads the bootloader into RMM memory during the platform boot process. 
When the platform boots, the trusted firmware allocates memory for the RMM, loads it and starts the RMM boot process. 
\codename extends this and allocates extra memory for the RMM in the realm and places the \tsfw into it. 
The trusted firmware always loads the \tsfw into a fixed physical address that the RMM is programmed with. 
When the platform software (the trusted firmware and RMM) boot, CCA adds their measurements to a platform attestation report. 
With \codename, the \tsfw is a part of the platform software and is included in CCA's platform attestation report. 
CCA includes this platform attestation report in all \sservice reports queried by remote verifiers (see~\cref{ssec:problem-formulation}). 
Further, the modularity of our design that separates \tsfw from the RMM ensures that phone manufacturers can update the \tsfw through platform updates to the device. 

During \sservice creation, the RMM copies the \tsfw into \sservice memory and sets it as the entry point for the \sservice boot. 
Because the \tsfw is now a part of the \sservice memory, CCA's attestation process will add its measurements to the \sservice's attestation report. 
This ensures that the \tsfw can be checked by remote verifiers. 

When the \sservice boots, the \tsfw can be configured to first perform local attestation using CCA's measurements for the \sservice.
\codename makes a conscious choice to make this process configurable in the \tsfw. 
Specifically, platform developers can choose how the local attestation is performed and the set of measurements that are acceptable. 
Similarly, the \tsfw can be configured to check for specific digital signatures and only then boot the \sservice. 
Importantly for \codename, the bootloader parses the device trees, finds all devices attached to the \sservice and their configuration, and checks it to ensure that only allowed virtio devices are attached. 
Once the \tsfw completes the validation checks, it boots the \sservice.

%% file: sections/08_evaluation.tex
\section{Evaluation}
We perform our experiments on an Intel Xeon Gen5 processor, with 32\,cores\,/\,64\,threads and 184\,GB RAM. 

\subsection{Experiment Platform}

\noindent{\bf Qemu.}
The \android platform has instructions and a build target to execute it on the Arm Fixed Virtual Platform (FVP)~\cite{fvp_android}. 
However, this target is not actively maintained and it is not straightforward to functionally deploy \android on the FVP. 
After patching AOSP v13 to build and run successfully on the FVP, we still needed $>$\,$12$ hours to boot \android without any changes. 
Therefore, the FVP setup is infeasible to develop \codename on or perform our experiments. 
In contrast, Linaro's QEMU support for Arm CCA performs much better and takes $\approx 5$ minutes to boot \android. 
Therefore, we choose to use QEMU v$8.20$-rc4 with support for Arm CCA to prototype \codename.
We can only obtain number of instructions as  performance measurement from the FVP as cycle-accurate simulation is not possible for Arm CCA hardware~\cite{shelter, wang2024cage, acai-usenix}. 
To get the same performance metric for our experiments with QEMU, we use a customized version of QEMU's $\tt{insn}$ instruction tracer.

\noindent{\bf Measurement Setup.}
We use a setup (\setupbase) which executes an unmodified \android in the normal world with an unmodified CCA platform (RMM and \tf).
In this setup, we boot a Microdroid VM in the normal world without any protections using \android kernel's KVM as the hypervisor. 
We use \setupbase as our baseline to compare our setup (\setupour) against. 
For \setupour, we execute a CCA platform with \codename's changes to the \tf and the RMM. In the normal world, we execute \android on our modified Linux kernel with CCA patches. Then, we boot a realm VM as the \sservice with \codename protections. 
For both setups we run the VMs with $1$\,core and 1\,GB memory.

\subsection{Cost Breakdown} 
We run our experiments with the $2$ setups, \setupbase and \setupour and report the cost of \codename's security.

\noindent{\bf Platform Boot Costs.}
We measure the cost of booting the \tf, RMM and \android for our setups (see~\cref{tab:boot-costs}). 
We see a total overhead of $10.4\%$ during boot of \setupour. This slowdown is acceptable as this is a one-time cost (see~\cref{tab:boot-costs}). 
This overhead is because of the extra GPT that the \tf creates and populates during platform boot.
Acai uses a 2 GPT setup for prototyping and measures the overheads for setting up 2 GPTs on a Zynq UltraScale+ MPSoC ZCU102 with an Arm Cortex-A53 64-bit quad-core processor. 
\codename also needs to setup and populate 2 GPTs and would incur $20.8\%$ overhead to setup the additional GPT as a one-time cost during the platform boot (c.f., Acai~\cite{acai-usenix}).

\begin{table}[]
\centering
\caption{Host and guest boot costs in million \#instructions}
\label{tab:boot-costs}
\resizebox{0.8\columnwidth}{!}{%
\begin{tabular}{@{}lS[table-format=3.4]S[table-format=3.4]S[table-format=2.3]@{}}
\toprule
\textbf{Stage} & \textbf{\setupbase} & \textbf{\setupour} & \textbf{\%Overhead} \\ \midrule
\multicolumn{4}{c}{\bf \em Host}                                         \\ \midrule
BL1                       & 0.312                         & 0.316                        & 1.26                           \\
BL2                       & 42.891                       & 45.576                      & 6.26                          \\
BL31                      & 82.947                       & 82.954                      & 0.01                           \\
Kernel                    & 814.797                      & 918.598                     & 12.74                          \\ \midrule
\multicolumn{4}{c}{\bf \em Guest}                                        \\ \midrule
BL UBoot                & 1492.771                     & 4057.346                    & 171.80      \\
Kernel                    & 480.307                      & 4889.900                    & 918.08      \\
\bottomrule
\end{tabular}%
}
\end{table}

\begin{table*}[]
\centering
\caption{Case-studies Cost Breakdown. Number of context switches (columns 2-4), hypervisor-VM calls (columns 5-7), SMC (column 8), RMIs (column 9), RSIs (column 10), and number of instructions (in millions) for launching  (columns 11-13) and executing each application (columns 14-16) for baseline (\setupbase) and \codename (\setupour).}
\label{tab:benchmarks}
\resizebox{2\columnwidth}{!}{%
\begin{tabular}{@{}lrrrrrrrrrrrrrrr@{}}
\toprule
\multicolumn{1}{c}{\multirow{2}{*}{\textbf{\begin{tabular}[c]{@{}c@{}}App\\ No.\end{tabular}}}} &
  \multicolumn{3}{l}{\textbf{No. of Context Switches}} &
  \multicolumn{3}{l}{\textbf{Hypervisor to/from VM Calls}} &
  \multicolumn{3}{l}{\textbf{New Interface Calls in \setupour}} &
  \multicolumn{3}{l}{\textbf{Application Launch (Microdroid)}} &
  \multicolumn{3}{l}{\textbf{Application Execution (Payload)}} \\ \cmidrule(l){2-16} 
\multicolumn{1}{c}{} &
  \multicolumn{1}{c}{\setupbase} &
  \multicolumn{1}{c}{\setupour} &
  \multicolumn{1}{c}{\% Ovh} &
  \multicolumn{1}{c}{\setupbase} &
  \multicolumn{1}{c}{\setupour} &
  \multicolumn{1}{c}{\% Ovh} &
  \multicolumn{1}{c}{SMCs} &
  \multicolumn{1}{c}{RMIs} &
  \multicolumn{1}{c}{RSIs} &
  \multicolumn{1}{c}{\setupbase} &
  \multicolumn{1}{c}{\setupour} &
  \multicolumn{1}{c}{\% Ovh} &
  \multicolumn{1}{c}{\setupbase} &
  \multicolumn{1}{c}{\setupour} &
  \multicolumn{1}{c}{\% Ovh} \\ \midrule
App 1 & 302 & 2990 & 890.06  & 130833 & 541566 & 313.93 & 482461 & 541566 & 92 & 1269.605 & 2932.008 & 130.94 & 1905.165 & 4101.628 & 115.29 \\
App 2 & 246 & 2304 & 836.58  & 103090 & 423445 & 310.75 & 364395 & 423445 & 92 & 1289.332 & 2425.258 & 88.10  & 1731.185 & 3225.788 & 86.33  \\
App 3 & 301 & 3460 & 1049.50 & 130549 & 624385 & 378.27 & 624385 & 565287 & 92 & 1287.338 & 2890.267 & 124.52 & 1772.187 & 4058.681 & 129.02 \\ \bottomrule
\end{tabular}%
}
\end{table*}

\noindent{\bf \SService Boot.}
We measure the costs to boot a VM in both our setups and report that \setupour adds $257.06\%$ overhead. 
It is because \setupour performs 
significantly more context switches than \setupbase, all of which are routed through the RMM and \tf (\cref{tab:boot-costs}). 
In~\cref{tab:benchmarks} we report the creation and runtime overheads for all the \sservices with Microdroid Apps. 
The overheads are because each hypervisor call from the bootloader and kernel needs to go through the RMM and the \tf. 
Although high, this is a one-time setup cost for the \sservice. %

\noindent{\bf GPT Updates.}
We measure the overheads of \codename's additional GPT  when pages are delegated by invoking $\tt{rmi\_granule\_delegate}$ from the hypervisor in our 2 setups. 
\setupour needs $10$ additional instructions for each GPT update~\cite{acai-usenix}. 
Acai reports an overhead of 0.7\% for updating the additional GPT with their measurements on the Zynq board, so \codename would incur similar overheads.

%% file: sections/09_case-studies.tex
\subsection{Case Studies: \android \& Protected VM Apps}
\label{sec:case-studies}
We demonstrate \codename's compatibility with \android apps and existing pVMs. 
For this we use two existing pVM apps and implement an OTP generator app like in previous works~\cite{brasser2019sanctuary}. 
For the existing pVM apps, we see that once we setup the \codename platform with support to run Microdroid, we can execute the apps without any changes or developer effort. 
We run the apps in both our setups, start our instruction tracer when the \android app starts, and report the creation and runtime overheads in~\cref{tab:benchmarks}.
We see that all case studies invoke the same number of RSI calls---they are not app-specific, but are invoked by Microdroid which is used by all case studies.

\noindent{\bf App1: Google's Protected Downloads.}
Google's Private Compute Services app provides several services to get Google's sensitive data (e.g., models, heuristics) from the cloud~\cite{google-pcs} which we tested on a Pixel 8 phone. 
We found that one of these services launches a pVM to generate a public-private key pair for \emph{protected downloads}.
In this setting, Google launches pVMs to generate the key pair and send the public key to an \android app. 
To generate this key-pair, the pVM uses a VM-specific secret that it derives using local attestation during its boot. 
We migrate this pVM and use it as an \sservice to demonstrate the compatibility of \codename with existing \android use-cases. 
To compare this \setupour, we run the same VM payload in \setupbase.
After \sservice starts, \setupour executes $86 \%$ more instructions than \setupbase. 
\setupour performs total $890 \%$ more context switches to create, boot, and run the \sservice.

Another use-case for Google's pVMs is isolated compilation. 
Isolated compilation uses a pVM to download updates from Google's servers and compile them in an isolated environment inaccessible to \android. 
While this use-case fits \codename's \sservices model, we were unable to get the isolated compilation running on our Pixel phone and therefore could not run it with \codename as well. 

\noindent{\bf App2: Microdroid Test App.}
The \android platform contains a test app that deploys a Microdroid VM as a pVM. 
The VM takes as input two numbers from the \android app, adds them, and sends the result back to the \android app.
We execute this Microdroid VM with both our setups (\cref{tab:benchmarks}).
In total, \setupour executes $86 \%$ more instructions and performs $837 \%$ more context switches.

\noindent{\bf App3: OTP Generator.}
We implement an OTP generator \android app that spawns a Microdroid VM using \android's APIs. 
The \android app allows a user to scan a QR Code on a website to obtain a registration secret from the OTP server. 
It then spawns a Microdroid VM and sends it this registration secret. 
Once the VM boots, the app requests an OTP. 
The VM computes the OTP with the initial registration secret and sends it to the app. 
As expected, we see that \setupour performs $1050 \%$ more context switches than \setupbase due to the switches between the realm to normal worlds for communication. 
During runtime, we see that \setupour executes $129 \%$ more instructions than \setupbase.

Our app assumes a trust-on-first use approach to transfer the key to the VM. 
Sanctuary implements a similar OTP mechanism that executes in a sandboxed normal world \ta~\cite{brasser2019sanctuary}.
It proposes an OTP protocol that assumes that the OTP server has the public key of the \ta.
The OTP server transmits the initial registration key encrypted using the \ta's public key.
This protocol eliminates the trust-on-first use assumption and replaces it with a pre-shared key. 
We can implement a similar protocol and leverage the attestation primitives of the \sservice to setup a secure channel between the \sservice and the OTP server to transmit the registration key.

\noindent{\bf Remark: Comparison to CCA baselines.}
In our evaluation we compare the performance of \codename to regular VM executions using Linux KVM. Although \codename incurs significant overheads for \sservice creation and boot we emphasize that these measurements include the cost of running a VM using CCA. As shown in prior works~\cite{guarantee-cca}, moving a VM to CCA typically incurs such high costs for setup.

%% file: sections/10_related-work.tex
\section{Related Work}
In~\cref{ssec:tz-attacks}, we discuss previous works that address TrustZone security to motivate the need for \codename. Here, we start with prior works that consider the security of CCA.

\noindent{\bf Rethinking the Android TEE Landscape.} 
\codename is motivated by the wide range of attacks in the Android TEE ecosystem, as we summarized in Section~\ref{ssec:tz-attacks}.
\codename not only protects \sservices but also rethinks the protection for Android and Secure-world. 
This holistic re-examination leads us to choose a design that is functionally  compatible with the existing landscape and retrofits security.
\codename's choice of VM abstraction is motivated by realistic deployment scenarios in the Android ecosystem. \codename considers a mobile platform with \android and deploys \sservices in the realm world that can range from trusted binaries to full VMs. Native Android support for pKVM, Microdroid, and real-world deployments of pVMs indicate that this abstraction is here to stay. 

Next, we compare \codename to prior works that also use multiple GPTs. 
In particular, Shelter advocates for small SApps in normal world which do not trust the RMM~\cite{shelter}. It intentionally limits expressiveness and cannot support VM abstractions, mainly because of low-TCB design choices.
Further, the need to remove the RMM from the TCB leads to the choice of one GPT per SApp in Shelter.
Instead, \codename only requires two GPTs and uses the RMM for intra-VM isolation. 
Lastly, prior CCA-based approaches including Shelter only aim to offer one-way isolation i.e., protect the confidential workloads from software adversaries (host OS, hypervisor, secure world). 
So they do not aim to achieve sandboxing as well as mutual isolation, which is a novel contribution in \codename.

\noindent{\bf CCA-based VMs.}
Samsung Islet, Huawei, and Linaro provide CCA implementation stacks to deploy realm VMs~\cite{islet, huawei-qemu, linaro-cca}, while other works have taken steps towards verifying the CCA trusted software stack~\cite{li2022design, fox2023verification, lienabling}. 
In \codename, we use Linaro's software stack because it is officially supported by Arm. 
Islet aims to achieve an end-to-end usage, where an Android app launches a realm VM on the phone to download models from a confidential VM running in the cloud. 
However, Islet implementation currently only supports Linux VMs. For Android, it does not define a concrete security model for the mobile platform, or provide implementations to run the realm VMs with \android.
More importantly, it does not reason about all four primitives as we do in \codename which stems from our motivation to secure the Android TEE ecosystem. 
Future works can adopt \codename to Islet to improve its security. 

\noindent{\bf Device Support.}
Arm allows CCA-enabled integrated devices to connect to realm VMs and with optional RME device assignment extension it can connect TEE-enabled accelerators to realm VMs.
Cage and Acai use CCA to connect integrated devices and TEE-accelerators to realm VMs respectively~\cite{wang2024cage, acai-usenix}, with the right hardware support they can be used with \codename.
Shelter, Cage, and Acai use multiple GPTs either to isolate the shelter apps in the normal world or to isolate devices-accesses to realm memory. 
\codename leverages GPTs to strengthen CCA security and guarantee mutual isolation between normal, realm, and secure worlds with two GPTs.

\noindent{\bf Arm TEEs for Mobile Devices.}
Several previous works and phone manufacturers execute trusted kernels in secure world EL1~\cite{optee, trusty, sun2015trustice, qsee}, build language-runtime support to build isolated trusted services~\cite{tlrsantos}, or build \ta{s} for specialised secure world computation~\cite{li2014droidvault}.
The lack of sandboxing with these approaches has been exploited by attackers to compromise mobile  security~\cite{sok-tz-vuln, hpe, boomerang}. 
Learning from this, \codename builds sandboxing as a key security primitive in its design. 
Like CCA, TrustZone supports attaching secure devices which several works have explored~\cite{lentz2018secloak, liu2012software, trustUI, park2021rushmore, strongbox, park2023safe, yao2023minimizing}.

\noindent{\bf Interface Security In Arm.}
BinderCracker and FANS propose fuzzing methods to detect Binder interface vulnerabilities~\cite{feng2016bindercracker, liu2020fans, bugiel2012towards, dietz2011quire} which can be directly applied to \codename's \sservices to strengthen interface security.
Several previous works have proposed black-box fuzzing techniques for trusted services~\cite{busch2023teezz, partemu} and static analysis to detect interface vulnerabilities~\cite{hpe}. 
The insights from these methods can be applied to \codename's interfaces.
Several prior works have demonstrated interface attacks on Intel SGX, AMD SEV-SNP, and Intel TDX~\cite{asyncshock, gameofthreads, taleoftwoworlds, smashex2021, InkTag, coin, alder2023pandora, checkoway2013iago, heckler-usenix, wesee-oakland}. 
In response to these interface attacks, prior works have employed fuzzing~\cite{sgx-fuzz, FuzzSGX} and symbolic execution~\cite{alder2023pandora, TeeRex, coin, symgx} to detect interface bugs in SGX enclaves. 
This history of interface bugs breaking enclave security highlights the need for \codename interfaces.

\noindent{\bf Sandboxing Principles.}
Sandboxing principles have been extensively studied and adopted for applications beyond the Arm platform~\cite{mccune2008flicker, mccune2010trustvisor}. 
Software-based fault isolation is a longstanding technique to isolate processes from each other~\cite{wahbe1993efficient}. 
This technique has been used by several works to build sandboxes~\cite{ erlingsson2006xfi,mccamant2005efficient,mccamant2006evaluating, yedidia2024lightweight}. 
Google Native Client (NaCl) applies SFI to different architectures using hardware and software-based techniques to build sandboxes~\cite{yee2010native, sehr2010adapting}.
Another widely-used approach to sandboxing is through language-based mechanisms such as those adopted by WebAssembly where the compiler adds dynamic checks on memory accesses to isolate processes~\cite{haas2017bringing, bhargavan2023foundations, johnson2023wave}. 
\codename builds \sservices using Arm CCA.

\noindent{\bf Beyond Android.}
\codename's security principles can be useful in cloud security. 
In \codename, we use the flexibility that GPTs provide to ensure the mutual isolation principle. 
The lack of this principle has opened several attacks in Intel SGX~\cite{van2022sok} and several works have proposed mechanisms to address this problem~\cite{van2022sok, ahmad2021chancel, seo2017sgx, aex-notify, van2019tale}. 
Keystone ensures mutual isolation for enclaves with RISC-V~\cite{keystone}. 
Future work can look into enforcing this principle for Intel TDX and AMD SEV VMs using microcode, hardware changes, or by leveraging privilege levels~\cite{ahmad2023veil}. 
Finally, \codename's interface security principles can be useful for other TEEs that deploy VMs in the cloud. 
In fact, the VM abstraction in the cloud might make it easier to secure these interfaces.

%% file: sections/11_conclusion.tex
\section{Conclusion}

We analyze the Android TEE security landscape to conclude that TrustZone cannot enforce principle of least privilege to protect sensitive execution on Android phones.
We present a new design, \codename, to addresses gaps in TrustZone and trusted hypervisor based solutions. We first show that Arm CCA does not solve the least privilege problem. We repurpose it to enable sandboxed execution of secure services while being isolated from Android, secure world, as well as other secure services. Then, our use of hardware-based attestation and conscious interface design ensures \codename does not suffer from the security gaps and challenges of existing TEEs for Android. 
Our \codename prototype shows its feasibility and  ability to support existing SBSes on Android while promising acceptable runtime overheads. 
We hope that \codename unlocks a new TEE design space for the Android ecosystem to not only address prior security gaps but also open up new application avenues by the virtue of its security. 

%% file: sections/12_acknowledgements.tex
\section*{Acknowledgments}
We thank Friederike Groschupp for her feedback. This work was supported (in part) by the ETH4D Humanitarian Action Challenge Grant and Zurich Information Security and Privacy Center (ZISC).

%% file: main.bbl
\begin{thebibliography}{10}
\providecommand{\url}[1]{#1}
\csname url@samestyle\endcsname
\providecommand{\newblock}{\relax}
\providecommand{\bibinfo}[2]{#2}
\providecommand{\BIBentrySTDinterwordspacing}{\spaceskip=0pt\relax}
\providecommand{\BIBentryALTinterwordstretchfactor}{4}
\providecommand{\BIBentryALTinterwordspacing}{\spaceskip=\fontdimen2\font plus
\BIBentryALTinterwordstretchfactor\fontdimen3\font minus \fontdimen4\font\relax}
\providecommand{\BIBforeignlanguage}[2]{{%
\expandafter\ifx\csname l@#1\endcsname\relax
\typeout{** WARNING: IEEEtran.bst: No hyphenation pattern has been}%
\typeout{** loaded for the language `#1'. Using the pattern for}%
\typeout{** the default language instead.}%
\else
\language=\csname l@#1\endcsname
\fi
#2}}
\providecommand{\BIBdecl}{\relax}
\BIBdecl

\bibitem{mobile-os-share-2024}
{Ahmed Sherif}, ``{Market share of mobile operating systems worldwide from 2009 to 2024, by quarter},'' May 2024, {(2024). Accessed: Jun 6, 2024. [Online]. Available: \url{https://www.statista.com/statistics/272698/global-market-share-held-by-mobile-operating-systems-since-2009/}}.

\bibitem{android-security-docs}
{Google}, ``{Android Security},'' {(2024). Accessed: Jun 6, 2024. [Online]. Available: \url{https://source.android.com/docs/security}}.

\bibitem{sok-tz-vuln}
D.~Cerdeira, N.~Santos, P.~Fonseca, and S.~Pinto, ``Sok: Understanding the prevailing security vulnerabilities in trustzone-assisted tee systems,'' in \emph{2020 IEEE Symposium on Security and Privacy (SP)}, 2020, pp. 1416--1432.

\bibitem{boomerang}
\BIBentryALTinterwordspacing
P.~Jiang, Q.~Wang, J.~Cheng, C.~Wang, L.~Xu, X.~Wang, Y.~Wu, X.~Li, and K.~Ren, ``Boomerang: {Metadata-Private} messaging under hardware trust,'' in \emph{20th USENIX Symposium on Networked Systems Design and Implementation (NSDI 23)}.\hskip 1em plus 0.5em minus 0.4em\relax Boston, MA: USENIX Association, Apr. 2023, pp. 877--899. [Online]. Available: \url{https://www.usenix.org/conference/nsdi23/presentation/jiang}
\BIBentrySTDinterwordspacing

\bibitem{avf}
{Android}, ``{AVF architecture},'' {(2024). Accessed: Jun 6, 2024. [Online]. Available: \url{https://source.android.com/docs/core/virtualization/architecture}}.

\bibitem{cca}
ARM, ``{Arm Confidential Compute Architecture ({ARM-CCA})},'' \url{https://www.arm.com/why-arm/architecture/security-features/arm-confidential-compute-architecture}.

\bibitem{ma2023return}
Z.~Ma, X.~Tan, L.~Ziarek, N.~Zhang, H.~Hu, and Z.~Zhao, ``Return-to-non-secure vulnerabilities on arm cortex-m trustzone: Attack and defense,'' in \emph{2023 60th ACM/IEEE Design Automation Conference (DAC)}.\hskip 1em plus 0.5em minus 0.4em\relax IEEE, 2023, pp. 1--6.

\bibitem{global-platform}
{GlobalPlatform}, ``{TEE Client API Specification v1.0 | GPD\_SPE\_007},'' {(2024). Accessed: Jun 6, 2024. [Online]. Available: \url{https://globalplatform.org/specs-library/tee-client-api-specification/}}.

\bibitem{chen2017downgrade}
Y.~Chen, Y.~Zhang, Z.~Wang, and T.~Wei, ``Downgrade attack on trustzone,'' \emph{arXiv preprint arXiv:1707.05082}, 2017.

\bibitem{brasser2019sanctuary}
F.~Brasser, D.~Gens, P.~Jauernig, A.-R. Sadeghi, and E.~Stapf, ``Sanctuary: Arming trustzone with user-space enclaves.'' in \emph{NDSS}.\hskip 1em plus 0.5em minus 0.4em\relax NDSS, 2019.

\bibitem{li2019teev}
W.~Li, Y.~Xia, L.~Lu, H.~Chen, and B.~Zang, ``Teev: Virtualizing trusted execution environments on mobile platforms,'' in \emph{Proceedings of the 15th ACM SIGPLAN/SIGOPS international conference on virtual execution environments}, 2019, pp. 2--16.

\bibitem{cerdeira2022rezone}
D.~Cerdeira, J.~Martins, N.~Santos, and S.~Pinto, ``{ReZone}: Disarming {TrustZone} with {TEE} privilege reduction,'' in \emph{31st USENIX Security Symposium (USENIX Security 22)}, 2022, pp. 2261--2279.

\bibitem{hua2017vtz}
Z.~Hua, J.~Gu, Y.~Xia, H.~Chen, B.~Zang, and H.~Guan, ``vtz: Virtualizing arm trustzone,'' in \emph{{USENIX} Security}, 2017.

\bibitem{gunyah}
{Qualcomm Innovation Center, Inc.}, ``{Gunyah Hypervisor},'' {(2024). Accessed: Jun 6, 2024. [Online]. Available: \url{https://github.com/quic/gunyah-hypervisor?tab=readme-ov-file}}.

\bibitem{li2021twinvisor}
D.~Li, Z.~Mi, Y.~Xia, B.~Zang, H.~Chen, and H.~Guan, ``Twinvisor: Hardware-isolated confidential virtual machines for arm,'' in \emph{Proceedings of the ACM SIGOPS 28th Symposium on Operating Systems Principles}, 2021, pp. 638--654.

\bibitem{hafnium}
{Google}, ``{Hafnium},'' {(2024). Accessed: Jun 6, 2024. [Online]. Available: \url{https://hafnium.googlesource.com/hafnium/+/HEAD/docs/Architecture.md}}.

\bibitem{kwon2019pros}
D.~Kwon, J.~Seo, Y.~Cho, B.~Lee, and Y.~Paek, ``Pros: Light-weight privatized se cure oses in arm trustzone,'' \emph{IEEE Transactions on Mobile Computing}, vol.~19, no.~6, pp. 1434--1447, 2019.

\bibitem{shelter}
\BIBentryALTinterwordspacing
Y.~Zhang, Y.~Hu, Z.~Ning, F.~Zhang, X.~Luo, H.~Huang, S.~Yan, and Z.~He, ``{SHELTER}: Extending arm {CCA} with isolation in user space,'' in \emph{32nd USENIX Security Symposium (USENIX Security 23)}.\hskip 1em plus 0.5em minus 0.4em\relax Anaheim, CA: USENIX Association, Aug. 2023, pp. 6257--6274. [Online]. Available: \url{https://www.usenix.org/conference/usenixsecurity23/presentation/zhang-yiming}
\BIBentrySTDinterwordspacing

\bibitem{elasticlave}
J.~Z. Yu, S.~Shinde, T.~E. Carlson, and P.~Saxena, ``Elasticlave: An efficient memory model for enclaves,'' in \emph{31st USENIX Security Symposium (USENIX Security 22)}, 2022, pp. 4111--4128.

\bibitem{feng2016bindercracker}
H.~Feng and K.~G. Shin, ``Bindercracker: Assessing the robustness of android system services,'' \emph{arXiv preprint arXiv:1604.06964}, 2016.

\bibitem{feng2016understanding}
------, ``Understanding and defending the binder attack surface in android,'' in \emph{Proceedings of the 32nd Annual Conference on Computer Security Applications}, 2016, pp. 398--409.

\bibitem{liu2020fans}
B.~Liu, C.~Zhang, G.~Gong, Y.~Zeng, H.~Ruan, and J.~Zhuge, ``$\{$FANS$\}$: Fuzzing android native system services via automated interface analysis,'' in \emph{29th USENIX Security Symposium (USENIX Security 20)}, 2020, pp. 307--323.

\bibitem{dynamic-trustzone}
{Jason Parker}, ``{Introducing Arm’s Dynamic TrustZone technology},'' {(2024). Accessed: Jun 6, 2024. [Online]. Available: \url{https://community.arm.com/arm-community-blogs/b/architectures-and-processors-blog/posts/introducing-arms-dynamic-trustzone-technology}}.

\bibitem{optee}
{Linaro}, ``{OP-TEE},'' {(2024). Accessed: Jun 6, 2024. [Online]. Available: \url{https://www.trustedfirmware.org/projects/op-tee/}}.

\bibitem{keystone}
D.~Lee, D.~Kohlbrenner, S.~Shinde, K.~Asanovi{\'c}, and D.~Song, ``Keystone: An open framework for architecting trusted execution environments,'' in \emph{EuroSys}, 2020.

\bibitem{wesee-oakland}
B.~Schlüter, S.~Sridhara, A.~Bertschi, and S.~Shinde, ``{WeSee: Using Malicious \#VC Interrupts to Break AMD SEV-SNP},'' in \emph{IEEE S\&P}, 2024.

\bibitem{heckler-usenix}
B.~Schlüter, S.~Sridhara, M.~Kuhne, A.~Bertschi, and S.~Shinde, ``{Heckler: Breaking Confidential VMs with Malicious Interrupts},'' in \emph{USENIX Security}, 2024.

\bibitem{arm-mec}
A.~Holdings, ``Arm architecture reference manual for a-profile architecture,'' \url{https://developer.arm.com/documentation/ddi0487/latest/}, 2023.

\bibitem{linaro-cca}
{Linaro}, ``{Building an RME stack for QEMU},'' {(2024). Accessed: Jun 5, 2024. [Online]. Available: \url{https://linaro.atlassian.net/wiki/pages/viewpage.action?pageId=29051027459&pageVersion=40}}.

\bibitem{rmm-linaro}
{ARM} and {Linaro}, ``{Realm Management Monitor for Qemu, v1.0-eac5},'' {(2023). Accessed: Jun 5, 2024. [Online]. Available: \url{https://git.codelinaro.org/linaro/dcap/rmm/-/tree/rmm-v1.0-eac5}}.

\bibitem{tf-linaro}
------, ``{Trusted Firmware for Qemu with CCA, v2.10},'' {(2023). Accessed: Jun 5, 2024. [Online]. Available: \url{https://git.codelinaro.org/linaro/dcap/tf-a/trusted-firmware-a/-/tree/v1.0-eac5}}.

\bibitem{linux-linaro}
{Linux} and {Linaro}, ``{Linux Kernel for Qemu with CCA, v6.7-rc4 },'' {(2023). Accessed: Jun 5, 2024. [Online]. Available: \url{https://gitlab.arm.com/linux-arm/linux-cca/-/tree/cca-full/rmm-v1.0-eac5}}.

\bibitem{aosp-13}
{Google}, ``{Manifest for Android 13.0.0 Release 12},'' {(2022). Accessed: Jun 5, 2024. [Online]. Available: \url{https://android.googlesource.com/platform/manifest/+/refs/heads/android-13.0.0_r12}}.

\bibitem{aosp-kernel-15}
------, ``{Manifest for Android Kernel 15-6.6},'' {(2024). Accessed: Jun 5, 2024. [Online]. Available: \url{https://android.googlesource.com/kernel/manifest/+/refs/heads/common-android15-6.6}}.

\bibitem{fvp_android}
{Android}, ``{Build and run an Android system image targeting the ARM Fixed Virtual Platform or QEMU.}'' {(2024). Accessed: Jun 6, 2024. [Online]. Available: \url{https://cs.android.com/android/platform/superproject/main/+/main:device/generic/goldfish/fvpbase/}}.

\bibitem{wang2024cage}
C.~Wang, F.~Zhang, Y.~Deng, K.~Leach, J.~Cao, Z.~Ning, S.~Yan, and Z.~He, ``Cage: Complementing arm cca with gpu extensions.''\hskip 1em plus 0.5em minus 0.4em\relax ISOC, 2024.

\bibitem{acai-usenix}
S.~Sridhara, A.~Bertschi, B.~Schl{\"u}ter, M.~Kuhne, F.~Aliberti, and S.~Shinde, ``{Acai: Protecting Accelerator Execution with Arm Confidential Computing Architecture},'' in \emph{USENIX Security}, 2024.

\bibitem{google-pcs}
{Google}, ``{Android Private Compute Services},'' {(2024). Accessed: Jun 6, 2024. [Online]. Available: \url{https://github.com/google/private-compute-services}}.

\bibitem{guarantee-cca}
\BIBentryALTinterwordspacing
S.~Siby, S.~Abdollahi, M.~Maheri, M.~Kogias, and H.~Haddadi, ``Guarantee: Towards attestable and private ml with cca,'' in \emph{Proceedings of the 4th Workshop on Machine Learning and Systems}, ser. EuroMLSys '24.\hskip 1em plus 0.5em minus 0.4em\relax New York, NY, USA: Association for Computing Machinery, 2024, p. 1–9. [Online]. Available: \url{https://doi.org/10.1145/3642970.3655845}
\BIBentrySTDinterwordspacing

\bibitem{islet}
{Samsung}, ``{Islet},'' {(2024). Accessed: Jun 6, 2024. [Online]. Available: \url{https://islet-project.github.io/islet/}}.

\bibitem{huawei-qemu}
{Huawei}, ``{Huawei\_CCA\_QEMU},'' {(2024). Accessed: Jun 6, 2024. [Online]. Available: \url{https://github.com/Huawei/Huawei_CCA_QEMU}}.

\bibitem{li2022design}
X.~Li, X.~Li, C.~Dall, R.~Gu, J.~Nieh, Y.~Sait, and G.~Stockwell, ``Design and verification of the arm confidential compute architecture,'' in \emph{16th USENIX Symposium on Operating Systems Design and Implementation (OSDI 22)}, 2022, pp. 465--484.

\bibitem{fox2023verification}
A.~C. Fox, G.~Stockwell, S.~Xiong, H.~Becker, D.~P. Mulligan, G.~Petri, and N.~Chong, ``A verification methodology for the arm{\textregistered} confidential computing architecture: From a secure specification to safe implementations,'' \emph{Proceedings of the ACM on Programming Languages}, vol.~7, no. OOPSLA1, pp. 376--405, 2023.

\bibitem{lienabling}
X.~Li, X.~Li, C.~Dall, R.~Gu, J.~Nieh, Y.~Sait, G.~Stockwell, M.~Knight, and C.~Garcia-Tobin, ``Enabling realms with the arm confidential compute architecture.''

\bibitem{trusty}
{Android}, ``{Trusty TEE},'' {(2024). Accessed: Jun 6, 2024. [Online]. Available: \url{https://source.android.com/docs/security/features/trusty}}.

\bibitem{sun2015trustice}
H.~Sun, K.~Sun, Y.~Wang, J.~Jing, and H.~Wang, ``Trustice: Hardware-assisted isolated computing environments on mobile devices,'' in \emph{2015 45th Annual IEEE/IFIP International Conference on Dependable Systems and Networks}.\hskip 1em plus 0.5em minus 0.4em\relax IEEE, 2015, pp. 367--378.

\bibitem{qsee}
{Qualcomm}, ``{Guard your data with the qualcomm snapdragon mobile platform},'' {(2024). Accessed: Jun 6, 2024. [Online]. Available: \url{https://www.qualcomm.com/content/dam/qcomm-martech/dm-assets/documents/guard_your_data_with_the_qualcomm_snapdragon_mobile_platform2.pdf}}.

\bibitem{tlrsantos}
\BIBentryALTinterwordspacing
N.~Santos, H.~Raj, S.~Saroiu, and A.~Wolman, ``Using arm trustzone to build a trusted language runtime for mobile applications,'' \emph{SIGARCH Comput. Archit. News}, vol.~42, no.~1, p. 67–80, feb 2014. [Online]. Available: \url{https://doi.org/10.1145/2654822.2541949}
\BIBentrySTDinterwordspacing

\bibitem{li2014droidvault}
X.~Li, H.~Hu, G.~Bai, Y.~Jia, Z.~Liang, and P.~Saxena, ``Droidvault: A trusted data vault for android devices,'' in \emph{2014 19th International Conference on Engineering of Complex Computer Systems}.\hskip 1em plus 0.5em minus 0.4em\relax IEEE, 2014, pp. 29--38.

\bibitem{hpe}
\BIBentryALTinterwordspacing
D.~Suciu, S.~McLaughlin, L.~Simon, and R.~Sion, ``Horizontal privilege escalation in trusted applications,'' in \emph{29th USENIX Security Symposium (USENIX Security 20)}.\hskip 1em plus 0.5em minus 0.4em\relax USENIX Association, Aug. 2020. [Online]. Available: \url{https://www.usenix.org/conference/usenixsecurity20/presentation/suciu}
\BIBentrySTDinterwordspacing

\bibitem{lentz2018secloak}
M.~Lentz, R.~Sen, P.~Druschel, and B.~Bhattacharjee, ``Secloak: Arm trustzone-based mobile peripheral control,'' in \emph{Proceedings of the 16th Annual International Conference on Mobile Systems, Applications, and Services}, 2018, pp. 1--13.

\bibitem{liu2012software}
H.~Liu, S.~Saroiu, A.~Wolman, and H.~Raj, ``Software abstractions for trusted sensors,'' in \emph{Proceedings of the 10th international conference on Mobile systems, applications, and services}, 2012, pp. 365--378.

\bibitem{trustUI}
W.~Li, M.~Ma, J.~Han, Y.~Xia, B.~Zang, C.-K. Chu, and T.~Li, ``Building trusted path on untrusted device drivers for mobile devices,'' in \emph{Proceedings of 5th Asia-Pacific Workshop on Systems}, ser. APSys ’14.\hskip 1em plus 0.5em minus 0.4em\relax New York, NY, USA: Association for Computing Machinery, 2014.

\bibitem{park2021rushmore}
C.~M. Park, D.~Kim, D.~V. Sidhwani, A.~Fuchs, A.~Paul, S.-J. Lee, K.~Dantu, and S.~Y. Ko, ``Rushmore: securely displaying static and animated images using trustzone,'' in \emph{Proceedings of the 19th Annual International Conference on Mobile Systems, Applications, and Services}, 2021, pp. 122--135.

\bibitem{strongbox}
Y.~Deng, C.~Wang, S.~Yu, S.~Liu, Z.~Ning, K.~Leach, J.~Li, S.~Yan, Z.~He, J.~Cao \emph{et~al.}, ``Strongbox: A gpu tee on arm endpoints,'' in \emph{Proceedings of the 2022 ACM SIGSAC Conference on Computer and Communications Security}, 2022, pp. 769--783.

\bibitem{park2023safe}
H.~Park and F.~X. Lin, ``Safe and practical gpu computation in trustzone,'' in \emph{Proceedings of the Eighteenth European Conference on Computer Systems}, 2023, pp. 505--520.

\bibitem{yao2023minimizing}
Z.~Yao, S.~M. Seyed~Talebi, M.~Chen, A.~Amiri~Sani, and T.~Anderson, ``Minimizing a smartphone's tcb for security-critical programs with exclusively-used, physically-isolated, statically-partitioned hardware,'' in \emph{Proceedings of the 21st Annual International Conference on Mobile Systems, Applications and Services}, 2023, pp. 233--246.

\bibitem{bugiel2012towards}
S.~Bugiel, L.~Davi, A.~Dmitrienko, T.~Fischer, A.-R. Sadeghi, and B.~Shastry, ``Towards taming privilege-escalation attacks on android.'' in \emph{NDSS}, vol.~17, 2012, p.~19.

\bibitem{dietz2011quire}
M.~Dietz, S.~Shekhar, Y.~Pisetsky, A.~Shu, and D.~S. Wallach, ``Quire: Lightweight provenance for smart phone operating systems,'' in \emph{USENIX security symposium}, vol.~31.\hskip 1em plus 0.5em minus 0.4em\relax San Francisco, CA;, 2011, p.~3.

\bibitem{busch2023teezz}
M.~Busch, A.~Machiry, C.~Spensky, G.~Vigna, C.~Kruegel, and M.~Payer, ``Teezz: Fuzzing trusted applications on cots android devices,'' in \emph{2023 IEEE Symposium on Security and Privacy (SP)}.\hskip 1em plus 0.5em minus 0.4em\relax IEEE, 2023, pp. 1204--1219.

\bibitem{partemu}
\BIBentryALTinterwordspacing
L.~Harrison, H.~Vijayakumar, R.~Padhye, K.~Sen, and M.~Grace, ``{PARTEMU}: Enabling dynamic analysis of {Real-World} {TrustZone} software using emulation,'' in \emph{29th USENIX Security Symposium (USENIX Security 20)}.\hskip 1em plus 0.5em minus 0.4em\relax USENIX Association, Aug. 2020, pp. 789--806. [Online]. Available: \url{https://www.usenix.org/conference/usenixsecurity20/presentation/harrison}
\BIBentrySTDinterwordspacing

\bibitem{asyncshock}
N.~Weichbrodt, A.~Kurmus, P.~Pietzuch, and R.~Kapitza, ``Asyncshock: Exploiting synchronisation bugs in intel sgx enclaves,'' in \emph{Computer Security--ESORICS 2016: 21st European Symposium on Research in Computer Security, Heraklion, Greece, September 26-30, 2016, Proceedings, Part I 21}.\hskip 1em plus 0.5em minus 0.4em\relax Springer, 2016, pp. 440--457.

\bibitem{gameofthreads}
\BIBentryALTinterwordspacing
J.~R. Sanchez~Vicarte, B.~Schreiber, R.~Paccagnella, and C.~W. Fletcher, ``Game of threads: Enabling asynchronous poisoning attacks,'' in \emph{Proceedings of the Twenty-Fifth International Conference on Architectural Support for Programming Languages and Operating Systems}, ser. ASPLOS '20.\hskip 1em plus 0.5em minus 0.4em\relax New York, NY, USA: Association for Computing Machinery, 2020, p. 35–52. [Online]. Available: \url{https://doi.org/10.1145/3373376.3378462}
\BIBentrySTDinterwordspacing

\bibitem{taleoftwoworlds}
\BIBentryALTinterwordspacing
J.~Van~Bulck, D.~Oswald, E.~Marin, A.~Aldoseri, F.~D. Garcia, and F.~Piessens, ``A tale of two worlds: Assessing the vulnerability of enclave shielding runtimes,'' in \emph{Proceedings of the 2019 ACM SIGSAC Conference on Computer and Communications Security}, ser. CCS '19.\hskip 1em plus 0.5em minus 0.4em\relax New York, NY, USA: Association for Computing Machinery, 2019, p. 1741–1758. [Online]. Available: \url{https://doi.org/10.1145/3319535.3363206}
\BIBentrySTDinterwordspacing

\bibitem{smashex2021}
J.~Cui, J.~Z. Yu, S.~Shinde, P.~Saxena, and Z.~Cai, ``Smashex: Smashing sgx enclaves using exceptions,'' in \emph{Proceedings of the 2021 ACM SIGSAC Conference on Computer and Communications Security}, ser. CCS '21, 2021.

\bibitem{InkTag}
O.~S. Hofmann, S.~Kim, A.~M. Dunn, M.~Z. Lee, and E.~Witchel, ``Inktag: Secure applications on an untrusted operating system,'' \emph{SIGPLAN Not.}, vol.~48, no.~4, p. 265–278, Mar. 2013.

\bibitem{coin}
\BIBentryALTinterwordspacing
M.~R. Khandaker, Y.~Cheng, Z.~Wang, and T.~Wei, ``Coin attacks: On insecurity of enclave untrusted interfaces in sgx,'' in \emph{Proceedings of the Twenty-Fifth International Conference on Architectural Support for Programming Languages and Operating Systems}, ser. ASPLOS '20.\hskip 1em plus 0.5em minus 0.4em\relax New York, NY, USA: Association for Computing Machinery, 2020, p. 971–985. [Online]. Available: \url{https://doi.org/10.1145/3373376.3378486}
\BIBentrySTDinterwordspacing

\bibitem{alder2023pandora}
F.~Alder, L.-A. Daniel, D.~Oswald, F.~Piessens, and J.~Van~Bulck, ``Pandora: Principled symbolic validation of intel sgx enclave runtimes.''

\bibitem{checkoway2013iago}
S.~Checkoway and H.~Shacham, ``Iago attacks: why the system call api is a bad untrusted rpc interface,'' \emph{ACM SIGARCH Computer Architecture News}, vol.~41, no.~1, pp. 253--264, 2013.

\bibitem{sgx-fuzz}
\BIBentryALTinterwordspacing
T.~Cloosters, J.~Willbold, T.~Holz, and L.~Davi, ``{SGXFuzz}: Efficiently synthesizing nested structures for {SGX} enclave fuzzing,'' in \emph{31st USENIX Security Symposium (USENIX Security 22)}.\hskip 1em plus 0.5em minus 0.4em\relax Boston, MA: USENIX Association, Aug. 2022, pp. 3147--3164. [Online]. Available: \url{https://www.usenix.org/conference/usenixsecurity22/presentation/cloosters}
\BIBentrySTDinterwordspacing

\bibitem{FuzzSGX}
A.~Khan, M.~Zou, K.~Kim, D.~Xu, A.~Bianchi, and D.~J. Tian, ``Fuzzing sgx enclaves via host program mutations,'' in \emph{2023 IEEE 8th European Symposium on Security and Privacy (EuroS\&P)}, 2023, pp. 472--488.

\bibitem{TeeRex}
\BIBentryALTinterwordspacing
T.~Cloosters, M.~Rodler, and L.~Davi, ``{TeeRex}: Discovery and exploitation of memory corruption vulnerabilities in {SGX} enclaves,'' in \emph{29th USENIX Security Symposium (USENIX Security 20)}.\hskip 1em plus 0.5em minus 0.4em\relax USENIX Association, Aug. 2020, pp. 841--858. [Online]. Available: \url{https://www.usenix.org/conference/usenixsecurity20/presentation/cloosters}
\BIBentrySTDinterwordspacing

\bibitem{symgx}
\BIBentryALTinterwordspacing
Y.~Wang, Z.~Zhang, N.~He, Z.~Zhong, S.~Guo, Q.~Bao, D.~Li, Y.~Guo, and X.~Chen, ``Symgx: Detecting cross-boundary pointer vulnerabilities of sgx applications via static symbolic execution,'' in \emph{Proceedings of the 2023 ACM SIGSAC Conference on Computer and Communications Security}, ser. CCS '23.\hskip 1em plus 0.5em minus 0.4em\relax New York, NY, USA: Association for Computing Machinery, 2023, p. 2710–2724. [Online]. Available: \url{https://doi.org/10.1145/3576915.3623213}
\BIBentrySTDinterwordspacing

\bibitem{mccune2008flicker}
J.~M. McCune, B.~J. Parno, A.~Perrig, M.~K. Reiter, and H.~Isozaki, ``Flicker: An execution infrastructure for tcb minimization,'' in \emph{Proceedings of the 3rd ACM SIGOPS/EuroSys European Conference on Computer Systems 2008}, 2008, pp. 315--328.

\bibitem{mccune2010trustvisor}
J.~M. McCune, Y.~Li, N.~Qu, Z.~Zhou, A.~Datta, V.~Gligor, and A.~Perrig, ``Trustvisor: Efficient tcb reduction and attestation,'' in \emph{2010 IEEE Symposium on Security and Privacy}.\hskip 1em plus 0.5em minus 0.4em\relax IEEE, 2010, pp. 143--158.

\bibitem{wahbe1993efficient}
R.~Wahbe, S.~Lucco, T.~E. Anderson, and S.~L. Graham, ``Efficient software-based fault isolation,'' in \emph{Proceedings of the fourteenth ACM symposium on Operating systems principles}, 1993, pp. 203--216.

\bibitem{erlingsson2006xfi}
U.~Erlingsson, M.~Abadi, M.~Vrable, M.~Budiu, and G.~C. Necula, ``Xfi: Software guards for system address spaces,'' in \emph{Proceedings of the 7th symposium on Operating systems design and implementation}, 2006, pp. 75--88.

\bibitem{mccamant2005efficient}
S.~McCamant and G.~Morrisett, ``Efficient, verifiable binary sandboxing for a cisc architecture,'' 2005.

\bibitem{mccamant2006evaluating}
------, ``Evaluating sfi for a cisc architecture.'' in \emph{USENIX Security Symposium}, vol.~10, 2006, pp. 209--224.

\bibitem{yedidia2024lightweight}
Z.~Yedidia, ``Lightweight fault isolation: Practical, efficient, and secure software sandboxing,'' in \emph{Proceedings of the 29th ACM International Conference on Architectural Support for Programming Languages and Operating Systems, Volume 2}, 2024, pp. 649--665.

\bibitem{yee2010native}
B.~Yee, D.~Sehr, G.~Dardyk, J.~B. Chen, R.~Muth, T.~Ormandy, S.~Okasaka, N.~Narula, and N.~Fullagar, ``Native client: A sandbox for portable, untrusted x86 native code,'' \emph{Communications of the ACM}, vol.~53, no.~1, pp. 91--99, 2010.

\bibitem{sehr2010adapting}
D.~Sehr, R.~Muth, C.~Biffle, V.~Khimenko, E.~Pasko, K.~Schimpf, B.~Yee, and B.~Chen, ``Adapting software fault isolation to contemporary $\{$CPU$\}$ architectures,'' in \emph{19th USENIX Security Symposium (USENIX Security 10)}, 2010.

\bibitem{haas2017bringing}
A.~Haas, A.~Rossberg, D.~L. Schuff, B.~L. Titzer, M.~Holman, D.~Gohman, L.~Wagner, A.~Zakai, and J.~Bastien, ``Bringing the web up to speed with webassembly,'' in \emph{Proceedings of the 38th ACM SIGPLAN Conference on Programming Language Design and Implementation}, 2017, pp. 185--200.

\bibitem{bhargavan2023foundations}
K.~Bhargavan, J.~Protzenko, A.~Rossberg, and D.~Stefan, ``Foundations of webassembly,'' 2023.

\bibitem{johnson2023wave}
E.~Johnson, E.~Laufer, Z.~Zhao, D.~Gohman, S.~Narayan, S.~Savage, D.~Stefan, and F.~Brown, ``Wave: a verifiably secure webassembly sandboxing runtime,'' in \emph{2023 IEEE Symposium on Security and Privacy (SP)}.\hskip 1em plus 0.5em minus 0.4em\relax IEEE, 2023, pp. 2940--2955.

\bibitem{van2022sok}
S.~van Schaik, A.~Seto, T.~Yurek, A.~Batori, B.~AlBassam, C.~Garman, D.~Genkin, A.~Miller, E.~Ronen, and Y.~Yarom, ``Sok: Sgx. fail: How stuff get exposed,'' 2022.

\bibitem{ahmad2021chancel}
A.~Ahmad, J.~Kim, J.~Seo, I.~Shin, P.~Fonseca, and B.~Lee, ``Chancel: Efficient multi-client isolation under adversarial programs.'' in \emph{NDSS}, 2021.

\bibitem{seo2017sgx}
J.~Seo, B.~Lee, S.~M. Kim, M.-W. Shih, I.~Shin, D.~Han, and T.~Kim, ``Sgx-shield: Enabling address space layout randomization for sgx programs.'' in \emph{NDSS}, 2017.

\bibitem{aex-notify}
\BIBentryALTinterwordspacing
S.~Constable, J.~V. Bulck, X.~Cheng, Y.~Xiao, C.~Xing, I.~Alexandrovich, T.~Kim, F.~Piessens, M.~Vij, and M.~Silberstein, ``{AEX-Notify}: Thwarting precise {Single-Stepping} attacks through interrupt awareness for intel {SGX} enclaves,'' in \emph{32nd USENIX Security Symposium (USENIX Security 23)}.\hskip 1em plus 0.5em minus 0.4em\relax Anaheim, CA: USENIX Association, Aug. 2023, pp. 4051--4068. [Online]. Available: \url{https://www.usenix.org/conference/usenixsecurity23/presentation/constable}
\BIBentrySTDinterwordspacing

\bibitem{van2019tale}
J.~Van~Bulck, D.~Oswald, E.~Marin, A.~Aldoseri, F.~D. Garcia, and F.~Piessens, ``A tale of two worlds: Assessing the vulnerability of enclave shielding runtimes,'' in \emph{Proceedings of the 2019 ACM SIGSAC Conference on Computer and Communications Security}, 2019, pp. 1741--1758.

\bibitem{ahmad2023veil}
A.~Ahmad, B.~Ou, C.~Liu, X.~Zhang, and P.~Fonseca, ``Veil: A protected services framework for confidential virtual machines,'' in \emph{Proceedings of the 28th ACM International Conference on Architectural Support for Programming Languages and Operating Systems, Volume 4}, 2023, pp. 378--393.

\end{thebibliography}
